\documentclass{jfm}

\usepackage{graphicx}
\usepackage{newtxtext}
\usepackage{newtxmath}
\usepackage{natbib}
\usepackage{subcaption}
\usepackage{hyperref}
\hypersetup{
    colorlinks = true,
    urlcolor   = blue,
    citecolor  = black,
}

\newcommand{\RomanNumeralCaps}[1]
\linenumbers

\title{Transition to chaos in a reduced-order model of a shear layer}

\author{Andr\'e V. G. Cavalieri\aff{1}
  \corresp{\email{andre@ita.br}},
  Erico L. Rempel\aff{2}
 \and Petr\^onio A. S. Nogueira\aff{3}}

\affiliation{\aff{1}Divis\~ao de Engenharia Aeroespacial, Instituto Tecnol\'ogico de Aeron\'autica, S\~ao Jos\'e dos Campos, SP, Brazil
\aff{2}Divis\~ao de Ci\^encias Fundamentais, Instituto Tecnol\'ogico de Aeron\'autica, S\~ao Jos\'e dos Campos, SP, Brazil
\aff{3}Department of Mechanical and Aerospace Engineering, Laboratory for Turbulence Research in Aerospace and Combustion, Monash University, Clayton, Australia}

\begin{document}
\maketitle

\begin{abstract}
{The present work studies the non-linear dynamics of a shear layer, driven by a body force and confined between parallel walls, a simplified setting to study transitional and turbulent shear layers. It was introduced by Nogueira \& Cavalieri (J. Fluid Mech. 907, A32, 2021), and is here studied using a reduced-order model based on a Galerkin projection of the Navier-Stokes system. By considering a confined shear layer with free-slip boundary conditions on the walls, periodic boundary conditions in streamwise and spanwise directions may be used, simplifying the system and enabling the use of methods of dynamical systems theory. A basis of eight modes is used in the Galerkin projection, representing the mean flow, Kelvin-Helmholtz vortices, rolls, streaks and oblique waves, structures observed in the cited work, and also present in shear layers and jets. A dynamical system is obtained, and its transition to chaos is studied. Increasing Reynolds number $Re$ leads to pitchfork and Hopf bifurcations, and the latter leads to a limit cycle with amplitude modulation of vortices, as in the DNS by Nogueira \& Cavalieri. Further increase of $Re$ leads to the appearance of a chaotic saddle, followed by the emergence of quasi-periodic and chaotic attractors. The chaotic attractors suffer a merging crisis for higher $Re$, leading to chaotic dynamics with amplitude modulation and phase jumps of vortices. This is reminiscent of observations of coherent structures in turbulent jets, suggesting that the model represents dynamics consistent with features of shear layers and jets.}
\end{abstract}


{\bf MSC Codes }  {\it(Optional)} Please enter your MSC Codes here

\section{Introduction}

Coherent structures are an important feature in turbulent shear layers and jets, as recognised since the early 1970s \citep{crow1971orderly,brown1974density,moore1977role}. Large-scale turbulent structures are formed in shear layers with features similar to Kelvin-Helmholtz vortices observed in transitional flows. Their amplitude grows in space as structures are advected downstream, and as they reach a region with a thicker shear layer their amplitude saturates and then decays. Such behaviour may be modelled as an axially extended wavepacket, which is known to be a dominant source of jet noise \citep{jordan2013wave,cavalieri2019wave}. In turbulent jets, wavepackets are not dominant structures in terms of kinetic energy \citep{jaunet2017two}, but are nonetheless strongly correlated with the far-field sound~\citep{cavalieri2013wavepackets}. 

The low-frequency sound radiation by supersonic jets is related to Mach-wave radiation: the supersonic phase speed of coherent disturbances leads to radiated Mach waves~\citep{tam1995supersonic,jordan2013wave}. For subsonic jets the mechanism is more subtle. A hydrodynamic wave with subsonic phase speed does not radiate sound, and radiated acoustic waves are only obtained if one considers the amplitude modulation of wavepackets~\citep{crighton1975bpa,jordan2013wave}. However, recent studies have shown that the amplitude modulation is not sufficient to explain the far-field sound of subsonic jets, as modelling of disturbances as a time-periodic wavepacket leads to sound field more than 30dB lower than experimental observations~\citep{baqui2015coherence,breakey2017experimental}. The other relevant feature of wavepackets for subsonic noise generation is its jitter: wavepackets are aperiodic, with frequency-domain properties characterised by a decay of two-point coherence that greatly enhance its sound radiation~\citep{cavalieri2014coherence,baqui2015coherence,cavalieri2019wave}. Such aperiodic character leads to intermittent bursts in the acoustic radiation~\citep{hileman2005lss,bogey2007analysis,koenig2013farfield,kearney2013intermittent,akamine2019conditional}. Thus, to target jet noise, a proper model of coherent structures in subsonic jets should include both amplitude modulation and coherence decay.




In order to model wavepackets and their sound radiation it became standard to apply linearised models considering the jet mean field as a base flow. This follows from an early idea of \cite{crighton1976stability}, who modelled wavepackets by considering the equations of motion linearised around a slowly-diverging mean flow. Coherent structures are modelled as Kelvin-Helmholtz wavepackets excited at the nozzle exit. This leads to exponential amplification of disturbances near the nozzle, and downstream amplitude decay as the shear layer thickens. Predictions of such models lead to time-periodic wavepackets, with good agreement with the forced-jet experiments by \cite{crow1971orderly}. This was followed by other works, such as \cite{tam1984sound}, \cite{cohen1987evolution} and \cite{matsubara2020linear}, which confirmed the pertinence of linearised models for turbulent jets subject to harmonic excitation. The aforementioned predictions of time-periodic wavepackets were even seen to match results of unforced turbulent jets~\citep{suzuki2006instability,gudmundsson2011jfm,cavalieri2013wavepackets}, albeit with a perceivable mismatch in the downstream region.

However, besides amplitude modulation, it is important to understand the mechanisms of jitter in wavepackets. A time-periodic wavepacket is perfectly coherent by construction, and coherence decay can only be obtained through the introduction of stochastic disturbances. This is possible through the resolvent framework, with linearised equations forced by non-linear terms that are treated as external forcing. The framework relates inputs (non-linear forcing) and outputs (linearised flow responses), which are ranked based on gains (ratios of output and input energy). As a result, several response modes are obtained for a single frequency, and superposition of optimal and suboptimal response modes leads to stochastic behaviour, as reviewed by \cite{cavalieri2019wave}. 

Resolvent analysis for turbulence was initially developed for wall-bounded flows \citep{mckeon2010critical,hwang2010linear}, and resolvent models for jets were subsequently developed by \cite{garnaud2013preferred}, \cite{jeun2016input}. Recent comparisons between resolvent modes and numerical simulation or experiment were presented by \cite{schmidt2018spectral} and \cite{lesshafft2019resolvent}, showing good levels of agreement between the leading response mode and the dominant mode in the jet, obtained through spectral proper orthogonal decomposition \citep{towne2018spectral}. However, suboptimal modes do not show the same level of agreement, and hence resolvent analysis alone cannot provide a stochastic wavepacket sufficiently accurate for jet-noise predictions. Resolvent predictions are expected to match flow responses exactly if non-linear terms may be modelled as white noise~\citep{towne2018spectral,cavalieri2019wave}; this is not the case, and thus the inclusion of the structure of non-linear terms is relevant if one wishes to recover accurately the flow statistics, as recently exemplified in wall-bounded flows by \cite{nogueira2021forcing} and \cite{morra2021colour}.

Resolvent analysis sheds light on how non-linearities excite flow responses, but as non-linear terms are considered as an external forcing, further analysis is required to understand the intrinsic non-linear dynamics. Wall-bounded transitional and turbulent flows have seen significant advances by studies of non-linear dynamics. A first possibility is to obtain reduced-order models, where a low number of dominant coherent structures are considered in a Galerkin projection of the Navier-Stokes system. This has been pursued, for instance, by \cite{waleffe1997self}, \cite{moehlis2004low} and \cite{cavalieri2021structure}. The Galerkin projection leads to a low-order dynamical system, whose inspection shows the relevant non-linear interactions. The system may be examined using standard methods in nonlinear dynamics, and its low dimension allows fast computations of the response to various initial conditions, which allows a thorough study of the dynamics. The work of \cite{moehlis2004low} suggested that wall-bounded turbulence in small computational domains is transient, related to a chaotic saddle which retains the dynamics in a turbulent behaviour for long lifetimes, but eventually does not maintain chaos and the dynamics are attracted by the laminar solution. Such transient chaotic behaviour was later confirmed by experiments in pipe flow with controlled disturbances by \cite{hof2006finite}.

In more recent works, non-linear dynamics have been studied using the full resolution of direct numerical simulations (DNS) for low Reynolds number flows. For Couette flow, several non-trivial steady and periodic solutions have been found since the pioneering works of \cite{nagata1990three} and \cite{kawahara2001periodic}. There is now a large catalog of solutions, as reviewed by \cite{kawahara2012significance}. Such solutions are unstable, with a saddle behaviour in the state space. As solutions attract (repel) trajectories by their stable (unstable) manifolds, they provide structure to the state space, and it is possible to to understand chaotic features by proper ``maps'' of the state space including invariant solutions \citep{gibson2008visualizing}. More generally, chaotic dynamics comprise an infinity of unstable periodic solutions (or orbits), whose attracting and repelling properties may be related to the statistics of chaotic attractors via periodic-orbit theory \citep{cvitanovic1989periodic,chandler2013invariant,lucas2015recurrent}, although computations are challenging for complex flows.

The derivation of reduced-order models and the study of invariant solutions is simpler in wall-bounded flows due to their homogeneity in two directions (streamwise and spanwise, or azimuthal for pipe flow), which allows the use of periodic boundary conditions and small computational domains, referred to as minimal flow units \citep{jimenez1991minimal,hamilton1995regeneration,flores2010hierarchy}. A wavenumber decomposition greatly simplifies the analysis and modelling of wall-bounded flows. On the other hand, round jets have a single homogeneous direction, the azimuth, which makes even linear analysis more complex, requiring a global framework \citep{theofilis2003advances}. Accordingly, resolvent models are based on two-dimensional axisymmetric base flows, requiring a global solver with significant computational cost \citep{garnaud2013preferred,jeun2016input,schmidt2018spectral,lesshafft2019resolvent}. A further complication related to the non-homegeneity in streamwise direction is that jet dynamics and sound radiation are known to depend on the upstream boundary layer, as observed in experimental \citep{bridges1987roles,fontaine2015very} and numerical results \citep{bogey2010influence,bres2018importance}. A recent investigation by \cite{kaplan2020nozzle} suggests that turbulent-jet wavepackets are excited by coherent boundary-layer structures inside the nozzle. These would need to be included in a reduced-order model of turbulent jets. If one tries to circumvent this by imposing streamwise homogeneity, the result is a temporal jet, which leads to a different perspective in the analysis of dynamics and acoustics \citep{bogey2019noise}, but with a loss of temporal homogeneity, as the jet continually spreads in time. 

\cite{suponitsky2010linear} have carried out analysis of jet dynamics and noise by considering non-linear dynamics with a frozen base flow, maintained by a body force. This shows that a jet forced with two frequencies  $\omega_1$ and $\omega_2$ develops significant amplitudes in the difference frequency $\omega_2 - \omega_1$ in both hydrodynamic and acoustic fields, in agreement with the early experiments of \cite{ronneberger1979experiments}. \cite{wu2009low} explored a similar scenario, with the excitation of wavepackets with close frequencies but opposing azimuthal modes, leading to a slow mean flow distortion that radiates sound. More recently, \cite{wu2016nonlinear} and \cite{zhang2020nonlinear} derived a non-linear framework for a shear layer in the incompressible and compressible regimes, respectively, with initial disturbances introduced at the nozzle lip with a non-linear evolution in space and time. The effect of incoherent turbulence is accounted for by including a model of Reynolds stresses.

\begin{figure}
\centerline{\includegraphics[width=0.5\textwidth]{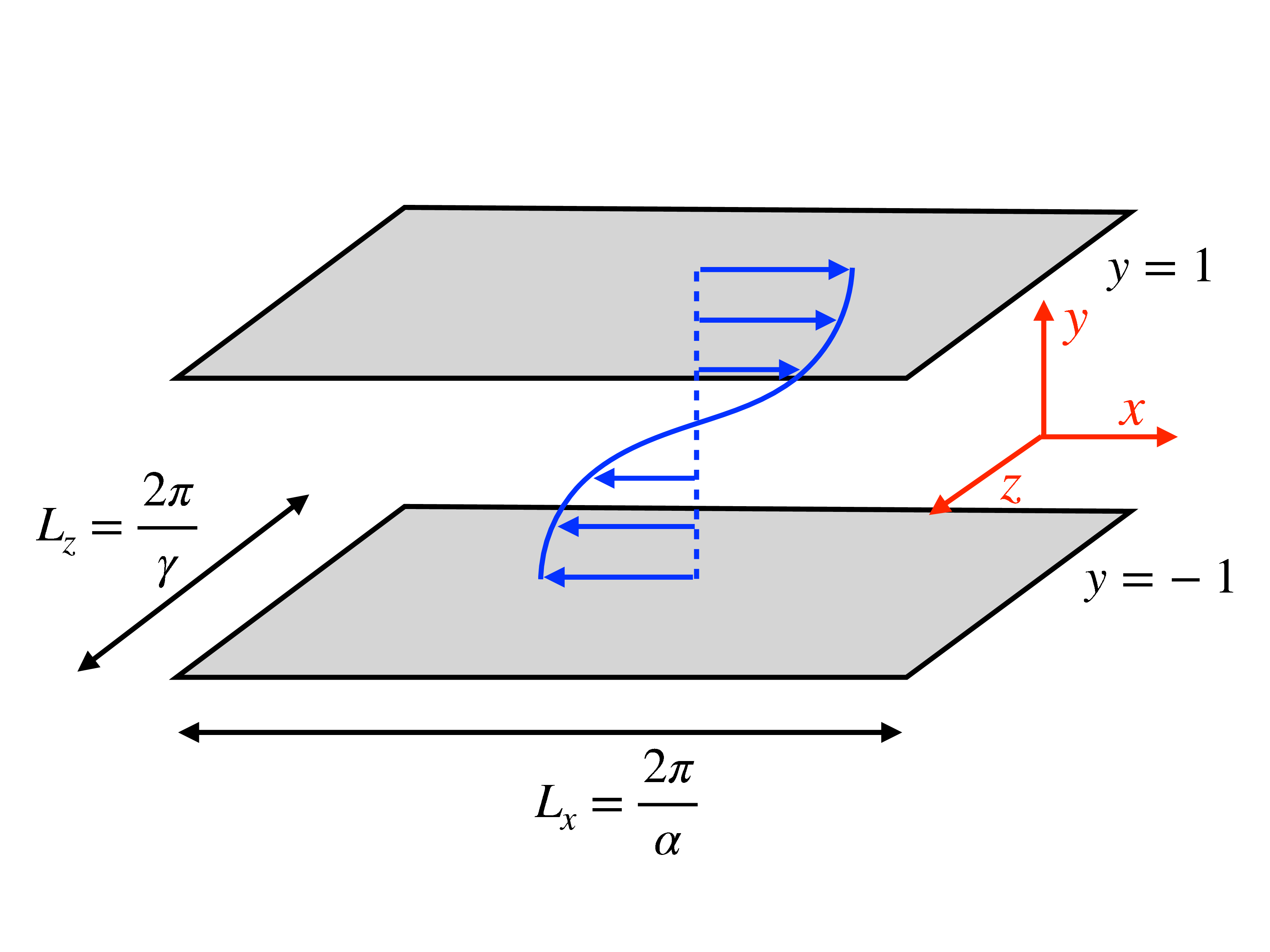}}
\caption{Sketch of the confined shear layer studied in NC (PAPU flow). Blue arrows show the applied body force (which leads to a similar velocity profile for the free-slip boundary conditions considered in this work).}
\label{fig:sketch}
\end{figure}

The above studies offer insight on non-linear interactions among wavepackets, with results that depend on the details of upstream excitation, a natural feature of shear layers and jets. In \cite{nogueira2021dynamics} (hereafter referred to as NC) we have adopted a different strategy and circumvented the upstream excitation problem by defining a shear layer that is homogeneous in the streamwise direction; this is accomplished by considering a shear layer driven by a body force and confined by two horizontal walls, as sketched in figure \ref{fig:sketch}. The body force leads to a velocity profile with an inflection point, leading to the Kelvin-Helmholtz instability typical of shear layers and jets, and homogeneity in streamwise and spanwise directions leads to a more straightforward extraction of coherent structures and analysis of their interplay. The effect of jet divergence is lost, which is beneficial for a dynamical system analysis as it allows defining a minimal flow unit for a shear layer; however, this comes with a cost, as the effect of mean-flow distortion in coherent structures appears in NC as a temporal modulation, instead of the spatial modulation of wavepackets seen in shear layers and jets \citep{gudmundsson2011jfm,cavalieri2013wavepackets,zhang2020nonlinear}.

It is seen in NC that the shear layer develops not only two-dimensional (spanwise constant) Kelvin-Helmholtz (KH) waves, but also oblique waves. KH waves are related to non-dimensional wavenumbers $(k_x, k_z) = (\pm1,0)$ in streamwise and spanwise direction, respectively; oblique waves are related to wavenumbers $(k_x, k_z) = (\pm1,\pm1)$. Moreover, the flow also displays streamwise vortices (or rolls) and streaks, with wavenumbers $(k_x, k_z) = (0,\pm1)$. These are known as dominant structures in wall-bounded turbulence~\citep{hamilton1995regeneration,del2006linear,hwang2010linear,abreu2020spectral}. Their presence in turbulent jets was recently studied by \cite{nogueira2019large} and \cite{pickering2020lift} by analyses of experimental and numerical databases, respectively, clarifying observations from earlier works \citep[see introduction by][for a detailed review and discussion]{pickering2020lift}. Visualisations from the velocity field in both cited works indicate that elongated streaks of streamwise velocity are the dominant structures in the velocity field at outer radial positions of turbulent jets, and such streaks are correlated to upstream streamwise vortices. 

The study in NC was based on a direct numerical simulation at a low Reynolds number of 200, and showed a limit cycle oscillation with an amplitude modulation in time of KH waves. This motivated labelling the configuration as permanently actuated, periodically unstable (PAPU) flow, as the continuous body force leads to a flow that display an alternation of KH instability and stability. The amplitude modulation was shown to be related to time-periodic mean flow distortion. Such distortion is related to the amplitude of rolls, streaks and oblique waves in the flow. These three structures undergo a periodic cycle. If one considers the sign convention in NC, positive oblique waves lead to the appearance of negative rolls, which lead to negative streaks by the lift-up effect, and finally to negative oblique waves by non-linear interactions, restarting the process with the opposite sign. The observations from the DNS motivated the proposition of an \emph{ad hoc} non-linear system, similar in spirit to the model for wall-bounded turbulence by \cite{waleffe1995transition}. The dynamical system in NC reproduced qualitatively the features of the limit cycle, giving hope that additional modelling work may lead to further insight about the flow at hand. Work in such simplified configuration, enabling most of the approaches in dynamical systems theory and chaos, has the potential of shedding light on more complex turbulent shear layers and jets.


{In this work we continue the exploration of the streamwise homogeneous shear layer in NC. Instead of non-slip boundary conditions used in NC, we consider free-slip conditions on the walls. This was motivated by the observation that increasing the Reynolds number in the NC configuration led to the appearance of near-wall turbulence, similar to what happens in turbulent Couette flow. This greatly departs from the desired parallel shear-layer behaviour, with walls that should act simply so as to prevent flow spreading. The use of free-slip boundary conditions avoids the occurence of such near-wall fluctuations, as shown in \cite{chantry2016turbulent}. Moreover, the use of free-slip boundary conditions enable deriving a low-order dynamical system using Fourier modes, following earlier works on wall-bounded turbulence \citep{waleffe1997self,moehlis2004low}. Unlike the \emph{ad hoc} dynamical system in NC, the present one is derived by a Galerkin projection of the Navier-Stokes equations. With such a simplified model the dynamics are nonetheless rich, and may be studied in some detail, with an exploration of the bifurcations of the system and its transition to chaotic behaviour, with features that are reminiscent of jitter seen in turbulent jets. The reduced-order model derived in the present work is thus based on the Navier-Stokes equations, and serves as a simplified setting that allows tracking the nonlinear dynamics of a shear layer, following the various bifurcations from the laminar solution towards chaotic dynamics as the Reynolds number is increased. Moreover, the solutions highlight mechanisms of interaction between streaks, vortices and oblique waves, leading to jitter in amplitude and phase of vortices. This opens new directions for investigation of more complex, spatially developing shear layers and jets.}

The remainder of the paper is organised as follows. Section~\ref{sec:ROMderivation} describes the procedure to obtain a reduced-order model and shows the resulting differential equations. Section \ref{sec:modelresults} presents an analysis of the non-linear dynamics of the model: initial bifurcations are studied in \ref{sec:bifurcations}, leading to a limit cycle similar to the one studied in NC. Further bifurcations of the system are studied in section \ref{sec:transitionchaos}. A symmetry-breaking bifurcation is shown in \ref{sec:symbreak}, the formation of a chaotic saddle is shown in \ref{sec:chaoticsaddle}, the transition to a chaotic attractor is explored in \ref{sec:finalchaos}, and an attractor merging crisis, leading to intermittent phase jumps in KH vortices, is studied in \ref{sec:mergingcrisis}. The work is completed with conclusions in section \ref{sec:conclusions}.


\section{Derivation of a reduced-order model}
\label{sec:ROMderivation}

We consider the incompressible flow between two parallel plates, driven by a body force, with a free-slip boundary condition on both walls. Non-dimensional quantities are obtained considering the half distance between walls and a reference velocity obtained from the laminar solution with sinusoidal forcing, following~\cite{moehlis2004low}. The coordinate system is Cartesian, with $x$, $y$ and $z$ referring respectively to streamwise, wall-normal and spanwise coordinates, as shown in figure \ref{fig:sketch}. Time is denoted by $t$.

We consider the system to be forced by a body force in the streamwise direction,
\begin{equation}
\mathbf{f} = 
\frac{\sqrt{2}}{\mathrm{Re}}
\left(\begin{array}{c} \beta ^2\,\sin\left(\beta \,y\right)+9\,P\,\beta ^2\,\sin\left(3\,\beta \,y\right)\\ 0\\ 0 \end{array}\right)
\end{equation}
with $\beta = \pi/2$, and $\mathrm{Re}$ being the Reynolds number.  With $P=0$ this forcing is equivalent to the one in sinusoidal shear flow (Wallefe flow) studied by~\cite{waleffe1997self} and~\cite{moehlis2004low}, and, more recently, by \cite{chantry2016turbulent,chantry2017universal} and \cite{cavalieri2021structure}. The cited works used this sinusoidal shear flow as a simpler system that retains features of wall-bounded turbulent flows, with chaotic behaviour despite the linear stability of the laminar solution for all Reynolds numbers. Here, the introduction of the PAPU parameter $P>0$ destabilises the laminar fixed point.

The flow so defined has a laminar fixed point given by
\begin{equation}
\mathbf{u}_L = 
\sqrt{2}\left(\begin{array}{c} \sin\left(\beta \,y\right)+9\,P\,\sin\left(3\,\beta \,y\right)\\ 0\\ 0 \end{array}\right)
\end{equation}
and thus the reference velocity is taken as $\sqrt{2}/2$ times the ($P=0$) laminar solution on the upper wall, $y=1$. This choice is the same of~\cite{moehlis2004low}, and will not be changed for $P\ne 0$ due to convenient properties of the reduced-order model that will be derived here. The present choice of body force is different from the $\tanh$ function considered in NC. Here the advantage is that the body force leads to a laminar solution that may be represented exactly with only two Fourier modes, which considerably simplifies the reduced-order model. The choice of the present body force leads  to a confined shear layer with an inviscid instability mechanism, similarly to NC.

We consider periodic boundary conditions in the streamwise and spanwise directions $x$ and $z$. The computational domain has dimensions $(L_x,L_y,L_z)=(2\pi/\alpha, 2, 2\pi / \gamma)$, where $\alpha$ and $\gamma$ are fundamental wavenumbers in $x$ and $z$ directions, respectively. 

We define an inner product as
\begin{equation}
\langle \mathbf{f}, \mathbf{g} \rangle = 
\frac{1}{2L_x L_z} \iiint{(f_x g_x + f_yg_y + f_zg_z) \mathrm{d}x\mathrm{d}y\mathrm{d}z}
\end{equation}
to use in a Galerkin projection of the Navier-Stokes equation. The use of free-slip and periodic boundary conditions leads to a natural use of Fourier modes to represent the velocity field, as sines or cosines with wavenumbers that are integer multiples of $\alpha$, $\beta$ and $\gamma$ satisfy the boundary conditions by construction. To model the structures observed in NC, we consider a set of eight orthonormal modes, given by
\begin{subequations}
\begin{equation}
\mathbf{u}_1 = 
\left(\begin{array}{c} \sqrt{2}\,\sin\left(\beta \,y\right)\\ 0\\ 0 \end{array}\right)
\end{equation}
\begin{equation}
\mathbf{u}_2 = 
\left(\begin{array}{c} \sqrt{2}\,\sin\left(3\,\beta \,y\right)\\ 0\\ 0 \end{array}\right)
\end{equation}
\begin{equation}
\mathbf{u}_3 = 
\left(\begin{array}{c} \frac{2\,\beta \,\sin\left(\alpha \,x\right)\,\sin\left(\beta \,y\right)}{\sqrt{\alpha ^2+\beta ^2}}\\ \frac{2\,\alpha \,\cos\left(\alpha \,x\right)\,\cos\left(\beta \,y\right)}{\sqrt{\alpha ^2+\beta ^2}}\\ 0 \end{array}\right)
\end{equation}
\begin{equation}
\mathbf{u}_4 = 
\left(\begin{array}{c} \frac{4\,\beta \,\cos\left(\alpha \,x\right)\,\cos\left(2\,\beta \,y\right)}{\sqrt{\alpha ^2+4\,\beta ^2}}\\ \frac{2\,\alpha \,\sin\left(\alpha \,x\right)\,\sin\left(2\,\beta \,y\right)}{\sqrt{\alpha ^2+4\,\beta ^2}}\\ 0 \end{array}\right)
\end{equation}
\begin{equation}
\mathbf{u}_5 = 
\left(\begin{array}{c} 0\\ \frac{2\,\gamma \,\cos\left(\beta \,y\right)\,\sin\left(\gamma \,z\right)}{\sqrt{\beta ^2+\gamma ^2}}\\ -\frac{2\,\beta \,\cos\left(\gamma \,z\right)\,\sin\left(\beta \,y\right)}{\sqrt{\beta ^2+\gamma ^2}} \end{array}\right)
\end{equation}
\begin{equation}
\mathbf{u}_6 = 
\left(\begin{array}{c} -\sqrt{2}\,\sin\left(\gamma \,z\right)\\ 0\\ 0 \end{array}\right)
\end{equation}
\begin{equation}
\mathbf{u}_7 = 
\left(\begin{array}{c} \frac{2\,\gamma \,\sin\left(\alpha \,x\right)\,\sin\left(\gamma \,z\right)}{\sqrt{\alpha ^2+\gamma ^2}}\\ 0\\ \frac{2\,\alpha \,\cos\left(\alpha \,x\right)\,\cos\left(\gamma \,z\right)}{\sqrt{\alpha ^2+\gamma ^2}} \end{array}\right)
\end{equation}
\begin{equation}
\mathbf{u}_8 = 
\left(\begin{array}{c} \frac{2\,\sqrt{2}\,\gamma \,\cos\left(\alpha \,x\right)\,\sin\left(\beta \,y\right)\,\sin\left(\gamma \,z\right)}{\sqrt{\alpha ^2+\gamma ^2}}\\ 0\\ -\frac{2\,\sqrt{2}\,\alpha \,\cos\left(\gamma \,z\right)\,\sin\left(\alpha \,x\right)\,\sin\left(\beta \,y\right)}{\sqrt{\alpha ^2+\gamma ^2}} \end{array}\right).
\end{equation}
\end{subequations}
Such modes are divergence-free and satisfy a free-slip condition on the walls at $y=\pm 1$. Hence, any linear superposition of modes also satisfies the same conditions. The wavenumbers ($k_x,k_y,k_z)$, in streamwise, wall-normal and spanwise directions, respectively, and physical structures related to each mode are shown in table \ref{tab:modes}. As modes have sinusoidal dependence in all directions, they always comprise wavenumbers $\pm k_x$, $\pm k_y$ and $\pm k_z$.

\begin{table}
\begin{center}
\def~{\hphantom{0}}
\begin{tabular}{c| c| c| c| c}
Mode & $k_x$ & $k_y$ & $k_z$ & Structure \\
$\mathbf{u}_1$ & 0 & $\beta$ & 0 & Mean flow 1 \\
$\mathbf{u}_2$ & 0 & $3\beta$ & 0 & Mean flow 2 \\
$\mathbf{u}_3$ & $\alpha$ & $\beta$ & 0 & Vortex 1 \\
$\mathbf{u}_4$ & $\alpha$ & $2\beta$ & 0 & Vortex 2 \\
$\mathbf{u}_5$ & 0 & $\beta$ & $\gamma$ & Roll \\
$\mathbf{u}_6$ & 0 & 0 & $\gamma$ & Streak \\
$\mathbf{u}_7$ & $\alpha$ & 0 & $\gamma$ & Oblique 1 \\
$\mathbf{u}_8$ & $\alpha$ & $\beta$ & $\gamma$ & Oblique 2 \\ 
\end{tabular}
\end{center}
\caption{Modes in the Galerkin projection}
\label{tab:modes}
\end{table}

The choice of modes is motivated by the wish of modelling the instability of a shear layer leading to Kelvin-Helmholtz vortices, the lift-up mechanism with rolls leading to streaks, and the interactions among these structures. Modes 1 and 2 are chosen to represent the laminar solution and its eventual mean-flow distortion due to Reynolds stresses. Modes 3 and 4 are a minimal representation of two-dimensional Kelvin-Helmholtz vortices with streamwise wavenumber $k_x=\alpha$; when considering triadic interactions in the Navier-Stokes system, we notice that the wavenumber of mode 4 is the sum of wavenumbers of modes 1 and 3, and the wavenumber of mode 4 is the sum of $k_x$ and the difference of $k_y$ of modes 2 and 3. These two consistent triads were seen to lead to a Kelvin-Helmholtz type instability, as will be shown in section \ref{sec:bifurcations}. Modes describing streamwise independent rolls (mode 5) and streaks (mode 6) are the same ones from \cite{waleffe1997self} and \cite{cavalieri2021structure}, and \cite{moehlis2004low} uses a slightly modified streak mode including two Fourier modes in $y$. 

Finally, triadic interactions between 2D vortex modes 3 and 4, and streamwise independent modes 5 and 6 are only possible if oblique-wave modes were included. This led to the choice of modes 7 and 8. A further analysis of the results in NC showed that the oblique modes had low amplitudes of the wall-normal velocity, and thus only $x$ and $z$ components were included in  modes 7 and 8. These modes are also present in the reduced-order models of sinusoidal shear flow by \cite{waleffe1997self} and \cite{moehlis2004low}. Their zero wall-normal velocity ensures that the laminar solution is stable to such disturbances, which would correspond to Squire modes which are always stable~\citep{schmid2001stability}.

We write the velocity field as a superposition of these eight modes,
\begin{equation}
\mathbf{u}(x,y,z,t) = \sum_j {a_j(t) \mathbf{u}_j(x,y,z)}
\label{eq:modaldecomposition}
\end{equation}
insert the decomposition in the Navier-Stokes equation and take an inner product with $\mathbf{u}_i$, in a Galerkin projection as used by \cite{moehlis2004low} and \cite{cavalieri2021structure}. This leads to 
\begin{equation}
\frac{\mathrm{d} a_i}{\mathrm{d} t}
=
F_i
+ \frac{1}{\mathrm{Re}}\sum_j{L_{i,j}}a_j
+ \sum_j{\sum_k{Q_{i,j,k}}a_ja_k}
\end{equation}
where the coefficients are given by
\begin{equation}
F_i = \langle \mathbf{f},\mathbf{u_i} \rangle,
\end{equation}
\begin{equation}
L_{i,j} = \langle \nabla^2 \mathbf{u_j},\mathbf{u_i} \rangle,
\end{equation}
\begin{equation}
Q_{i,j,k} = -\langle (\mathbf{u_j} \cdot \nabla) \mathbf{u_k},\mathbf{u_i} \rangle.
\end{equation}
As the modes satisfy an incompressibility condition and periodic boundary conditions, the contribution of the pressure term vanishes. The reduced-order model so obtained is a system of eight ordinary differential equations, given by
\begin{subequations}
\begin{eqnarray}
\dot{a_1} = 
\frac{\beta \,\left(2\,\beta -2\,a_{1}\,\beta \right)}{2\,\mathrm{Re}}-\frac{\beta \,\left(\frac{2\,a_{5}\,a_{6}\,\gamma }{k_{\beta,\gamma}}-\frac{3\,\sqrt{2}\,a_{3}\,a_{4}\,\alpha \,\beta }{k_{\alpha,2\beta}\,k_{\alpha,\beta}}\right)}{2}
\end{eqnarray}
\begin{eqnarray}
\dot{a_2} = 
\frac{9\,\beta ^2\,\left(P-a_{2}\right)}{\mathrm{Re}}+\frac{3\,\sqrt{2}\,a_{3}\,a_{4}\,\alpha \,\beta ^2}{2\,k_{\alpha,2\beta}\,k_{\alpha,\beta}}
\end{eqnarray}
\begin{eqnarray}
\dot{a_3} = 
-\frac{a_{3}\,\left(\alpha ^2+\beta ^2\right)}{\mathrm{Re}}-\frac{\sqrt{2}\,a_{1}\,a_{4}\,\alpha \,\left(\alpha ^2+3\,\beta ^2\right)}{2\,k_{\alpha,2\beta}\,k_{\alpha,\beta}}-\frac{\sqrt{2}\,a_{2}\,a_{4}\,\alpha \,\left(\alpha ^2-5\,\beta ^2\right)}{2\,k_{\alpha,2\beta}\,k_{\alpha,\beta}}\nonumber \\
-\frac{2\,a_{6}\,a_{8}\,\alpha \,\beta \,\gamma }{k_{\alpha,\beta}\,k_{\alpha,\gamma}}-\frac{a_{5}\,a_{7}\,\gamma ^2\,\left(\alpha ^2-\beta ^2\right)}{k_{\alpha,\beta}\,k_{\alpha,\gamma}\,k_{\beta,\gamma}}
\end{eqnarray}
\begin{eqnarray}
\dot{a_4} = 
-\frac{a_{4}\,\left(\alpha ^2+4\,\beta ^2\right)}{\mathrm{Re}}+\frac{\sqrt{2}\,a_{1}\,a_{3}\,\alpha ^3}{2\,k_{\alpha,2\beta}\,k_{\alpha,\beta}}
+\frac{\sqrt{2}\,a_{2}\,a_{3}\,\alpha \,\left(\alpha ^2-8\,\beta ^2\right)}{2\,k_{\alpha,2\beta}\,k_{\alpha,\beta}}\nonumber \\
+\frac{\sqrt{2}\,a_{5}\,a_{8}\,\gamma ^2\,\left(\alpha ^2-4\,\beta ^2\right)}{2\,k_{\alpha,2\beta}\,k_{\alpha,\gamma}\,k_{\beta,\gamma}}
\end{eqnarray}
\begin{eqnarray}
\dot{a_5} = 
-\frac{a_{5}\,\left(\beta ^2+\gamma ^2\right)}{\mathrm{Re}}+\frac{\sqrt{2}\,a_{4}\,a_{8}\,\alpha ^2\,\left(\beta ^2-\gamma ^2\right)}{2\,k_{\alpha,2\beta}\,k_{\alpha,\gamma}\,k_{\beta,\gamma}}-\frac{a_{3}\,a_{7}\,\alpha ^2\,\left(\beta ^2-\gamma ^2\right)}{k_{\alpha,\beta}\,k_{\alpha,\gamma}\,k_{\beta,\gamma}}
\end{eqnarray}
\begin{eqnarray}
\dot{a_6} = 
-\frac{a_{6}\,\gamma ^2}{\mathrm{Re}}+\gamma \,\left(\frac{a_{1}\,a_{5}\,\beta }{k_{\beta,\gamma}}+\frac{a_{3}\,a_{8}\,\alpha \,\beta }{k_{\alpha,\beta}\,k_{\alpha,\gamma}}\right)
\end{eqnarray}
\begin{eqnarray}
\dot{a_7} = 
-\frac{a_{7}\,\left(\alpha ^2+\gamma ^2\right)}{\mathrm{Re}}+a_{1}\,a_{8}\,\alpha +\frac{a_{3}\,a_{5}\,\beta ^2\,\left(\alpha ^2-\gamma ^2\right)}{k_{\alpha,\beta}\,k_{\alpha,\gamma}\,k_{\beta,\gamma}}
\end{eqnarray}
\begin{eqnarray}
\dot{a_8} = 
-\frac{a_{8}\,\left(\alpha ^2+\beta ^2+\gamma ^2\right)}{\mathrm{Re}}+\frac{a_{3}\,a_{6}\,\alpha \,\beta \,\gamma }{k_{\alpha,\beta}\,k_{\alpha,\gamma}}-a_{1}\,a_{7}\,\alpha -\frac{\sqrt{2}\,a_{4}\,a_{5}\,\beta ^2\,\left(\alpha ^2-4\,\gamma ^2\right)}{2\,k_{\alpha,2\beta}\,k_{\alpha,\gamma}\,k_{\beta,\gamma}},
\end{eqnarray}
\label{eq:ROM}
\end{subequations}
with auxiliary wavenumbers $k_{\alpha,\beta} = \sqrt{\alpha^2+\beta^2}$, $k_{\alpha,2\beta} = \sqrt{\alpha^2+4\beta^2}$, $k_{\alpha,\gamma} = \sqrt{\alpha^2+\gamma^2}$ and $k_{\beta,\gamma} = \sqrt{\beta^2+\gamma^2}$.

As expected from the Navier-Stokes equation, the system above has quadratic terms that conserve energy, only distributing it among modes. The equation for the integrated kinetic energy, $E = 1/2\sum_{i=1}^8{a_i^2}$, is.

\begin{eqnarray}
\frac{\mathrm{d}E}{\mathrm{d}t}=
\frac{a_{1}\,\beta ^2+9\,P\,a_{2}\,\beta ^2}{2\,\mathrm{Re}} -\frac{1}{2\,\mathrm{Re}}\sum_i^8{\kappa_i a_i^2}
\end{eqnarray}
with $\kappa_i$ being the wavenumber of the $i$-th mode. The first term from the right-hand side represents production due to the applied body force, and the last terms are related to viscous dissipation.

 
 The above system admits the symmetries $(a_3,a_4,a_7,a_8) \to (-a_3,-a_4,-a_7,-a_8)$, which is a mirror symmetry following the $x$ axis (or equivalently an $L_x/2$ shift in the $x$ direction), and $(a_5,a_6,a_7,a_8) \to (-a_5,-a_6,-a_7,-a_8)$, a mirror symmetry following the $z$ axis (or equivalently an $L_z/2$ shift in the $z$ direction. The first group of modes has an $x$ dependence, and the last group has a $z$ dependence.
 
 An inspection of the reduced-order model (ROM) in eq. (\ref{eq:ROM}) shows that the mean shear is maintained by the body force. The laminar solution corresponds to $a_1 = 1$, $a_2 = P$, and $a_3=a_4=...=a_8=0$. A two-dimensional flow, given by modes 1-4, allows an energy transfer between mean-flow modes 1 and 2 and vortex modes 3 and 4, which will be seen as the mechanism for vortex growth by a Kelvin-Helmholtz (KH) instability. Consideration of a streamwise independent flow, given by modes 1, 2, 5 and 6, shows that there is also energy transfer between the mean-flow mode 1 and the streak mode 6, mediated by the roll mode 5; this is the well-known lift-up effect~\citep{ellingsen1975stability,brandt2014lift}. Finally, vortices, rolls and streaks have various interactions with the oblique waves 7 and 8.
 
\section{Model results}
\label{sec:modelresults}
 
Besides the Reynolds number $\mathrm{Re}$, the model  in eq. (\ref{eq:ROM}) has as parameters the PAPU body force constant $P$, and the streamwise and spanwise domain lengths $L_x$ and $L_z$ that define fundamental wavenumbers $\alpha$ and $\gamma$, respectively. In this work we will focus on $L_x = 4\pi$ ($\alpha = 1/2$) and $L_z=2\pi$ ($\gamma = 1$) as this was the domain size in NC, and also one of the domains considered by \cite{moehlis2004low} and \cite{cavalieri2021structure}. An exploration of model results with this domain size showed that $P=0.08$ led to dynamics similar to the observations in NC. In what follows we will explore the results of the model with $P=0.08$. A variation of $P$ will prove useful to understand the appearance of a chaotic saddle.

Numerical integration of eq. (\ref{eq:ROM}) was carried out using a standard 4th-5th order Runge-Kutta method. Unless otherwise specified, a random initial condition was used.
 
\subsection{Pitchfork and Hopf bifurcations}
\label{sec:bifurcations}
 

We have tracked fixed points of the system with the aforementioned choice of parameters, as a function of Reynolds number, to determine a bifurcation system. Fixed points $\tilde{\mathbf{a}}$ may be sought by looking for solutions of
\begin{equation}
F_i
+ \sum_j{L_{i,j}}\tilde{a}_j
+ \sum_j{\sum_k{Q_{i,j,k}}\tilde{a}_j\tilde{a}_k}=0,
\end{equation}
and their stability may be studied by the eigenvalue problem
\begin{equation}
\sum_j{L_{i,j}}a'_j
+ \sum_j{ \sum_k{ \left( Q_{i,j,k} + Q_{i,k,j}\right)\tilde{a}_k}}   a'_j
=
\sigma a'_i
\end{equation}
where $\mathbf{a}'$ are small disturbances to the fixed point $\tilde{\mathbf{a}}$. The system above implies a time dependence given by $\exp(\sigma_i t)$; accordingly, an eigenvalue $\sigma$ with positive real part indicates an instability.

\begin{figure}
\begin{subfigure}{0.5\textwidth}{\includegraphics[width=1.0\textwidth]{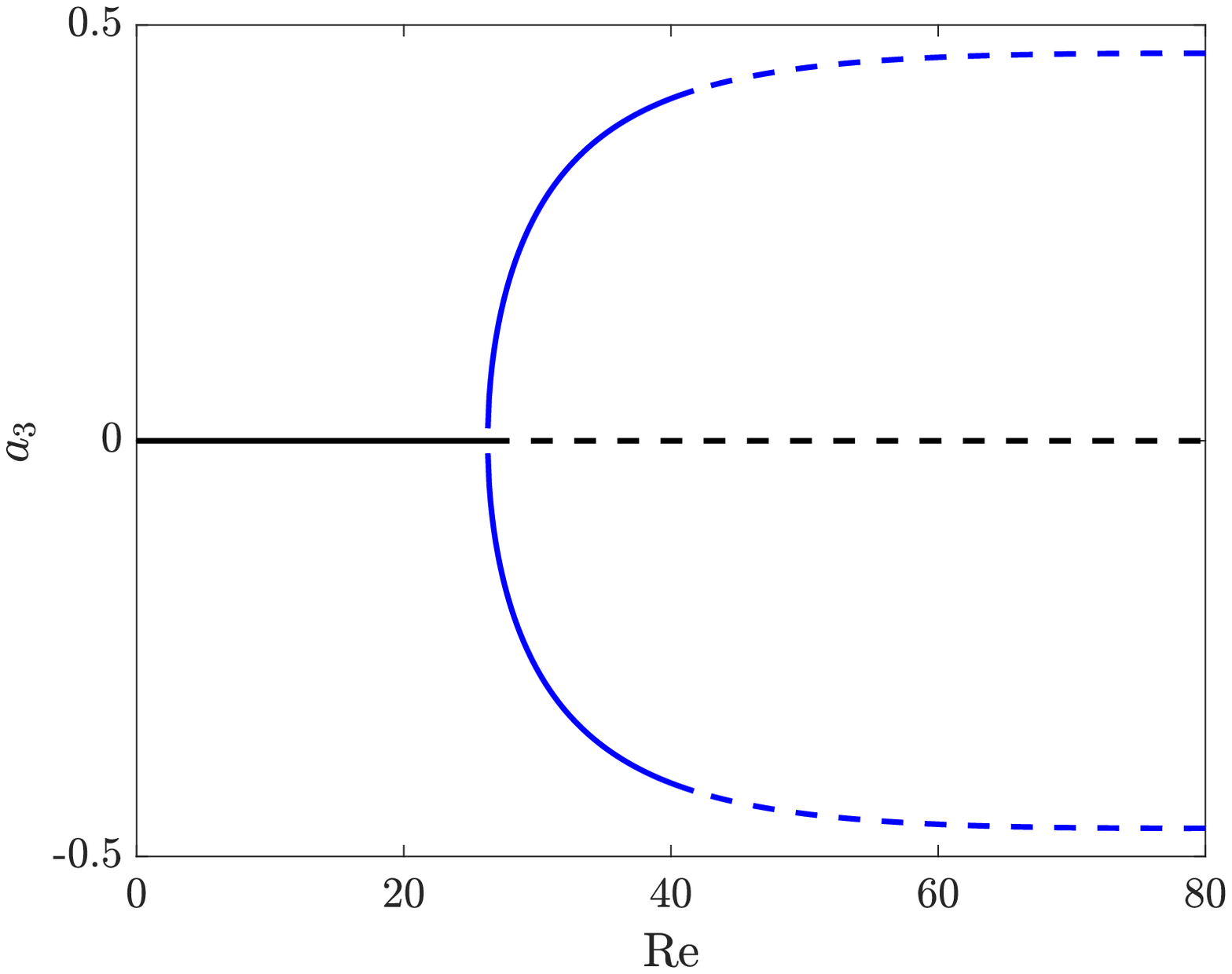}}\caption{Bifurcation diagram}\end{subfigure}\begin{subfigure}{0.5\textwidth}{\includegraphics[width=1.0\textwidth]{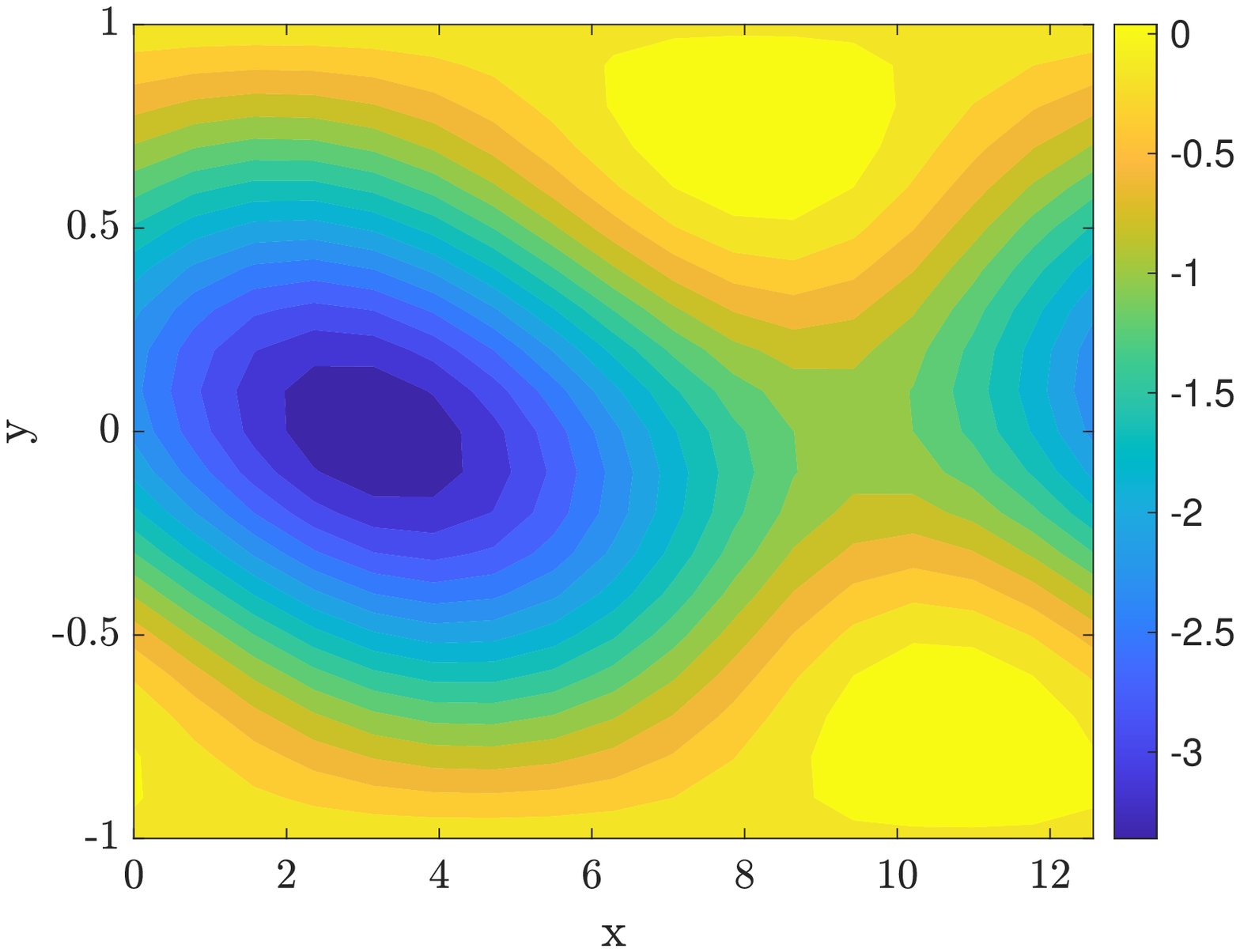}}\caption{$z$-vorticity of vortex solution for $\mathrm{Re}=35$}\end{subfigure}
\caption{Left: first pitchfork bifurcation of the system. Full lines are stable fixed points, and dashed lines show unstable ones. The black line refers to the laminar solution, and the blue line shows the saturated vortex. Right: $z$-vorticity of the saturated vortex solution for $\mathrm{Re}=35$.}
\label{fig:vortex_pitchfork}
\end{figure}

For $P=0.08$, the laminar solution in the ROM is stable up to $\mathrm{Re} = 26.3$. For Reynolds numbers larger than this value there is one real positive eigenvalue, indicating the occurrence of a pitchfork bifurcation. This was confirmed to be a supercritical bifurcation by a search for fixed points, whose results are shown in figure \ref{fig:vortex_pitchfork}. The new fixed points are saturated 2D vortex solutions, with non-zero amplitudes for the first four modes (the two-dimensional ones), and zero amplitudes for the remaining spanwise dependent modes. As the laminar solution and the body force are antisymmetric, these Kelvin-Helmholtz vortices have zero phase velocity and are thus a steady-state solution of the system after the pitchfork bifurcation. The two solutions differ only by an $L_x/2$ shift in the streamwise direction, which is expected due to the symmetry of the system. In what follows, we will consider these two solutions as ``positive'' or ``negative'' saturated vortex solutions based on the sign of $a_3$.

These saturated vortex modes are in turn stable up to $\mathrm{Re}=41.8$. A pair of complex-conjugate eigenvalues become unstable for Reynolds number beyond this value. By integrating the non-linear ROM in eq. (\ref{eq:ROM}) we verified that this is a supercritical Hopf bifurcation: a low-amplitude limit cycle appears for $\mathrm{Re}$ slightly larger than the critical value, and grows monotonically in amplitude for larger $\mathrm{Re}$. The emerging limit cycle involves all modes, and thus leads to three-dimensional behaviour. This may be related to the instabilities of shear-layer vortices studied by \cite{pierrehumbert1982two}, who have found that a periodic array of vortices has three dimensional instabilities which may be either stationary (referred to as translational instability) or oscillatory (higher-order modes). The stationary mode is studied by \cite{pierrehumbert1982two} as leading to the emergence of steady streaks, and the oscillatory modes have a more complex structure. As the present limit cycle involves all modes, which oscillate at a non-zero frequency, we tentatively associate the observed instability with the higher-order modes by \cite{pierrehumbert1982two}.

\begin{figure}
\begin{subfigure}{0.5\textwidth}{\includegraphics[width=1.0\textwidth]{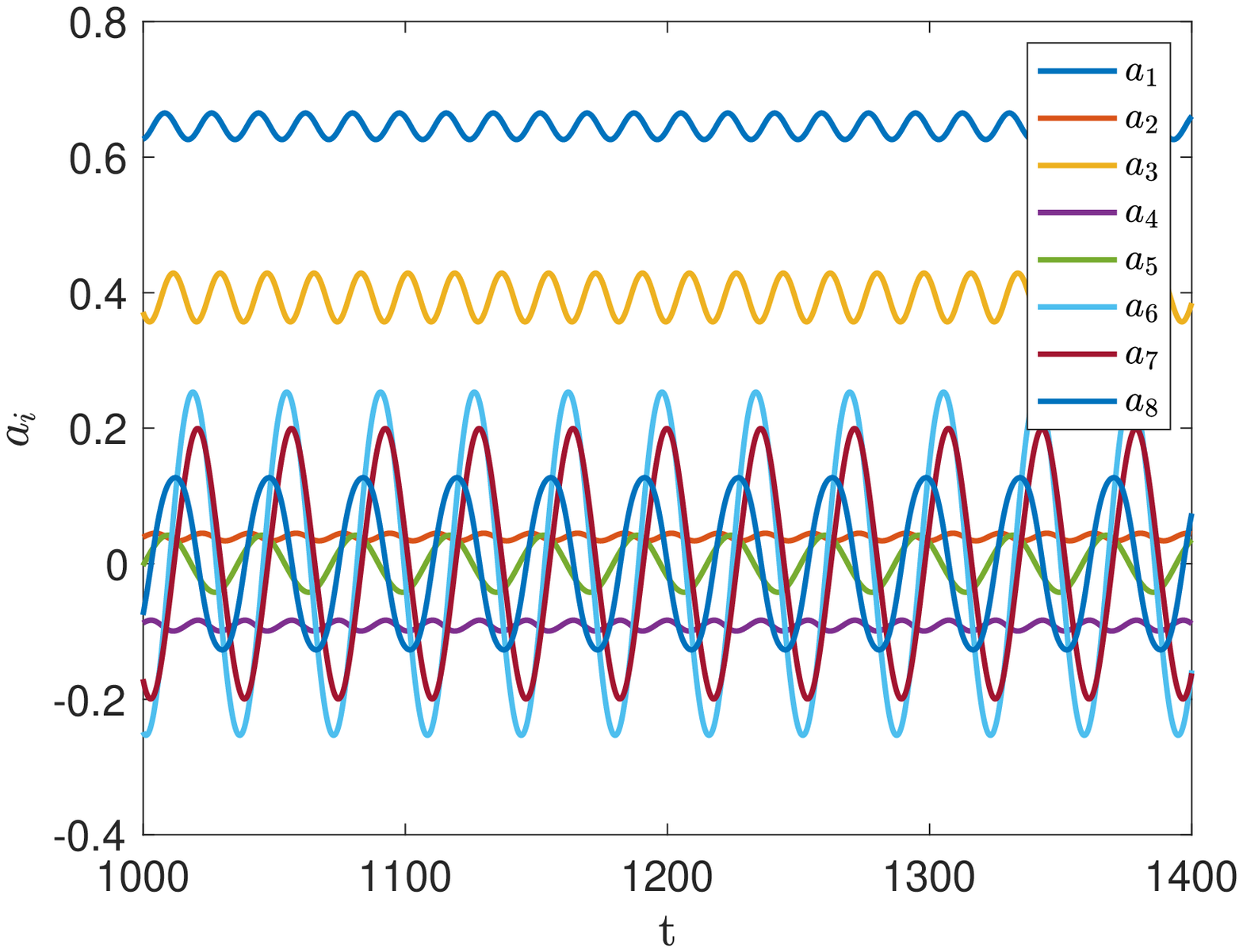}\caption{Sample time series}}\end{subfigure}
\begin{subfigure}{0.5\textwidth}{\includegraphics[width=1.0\textwidth]{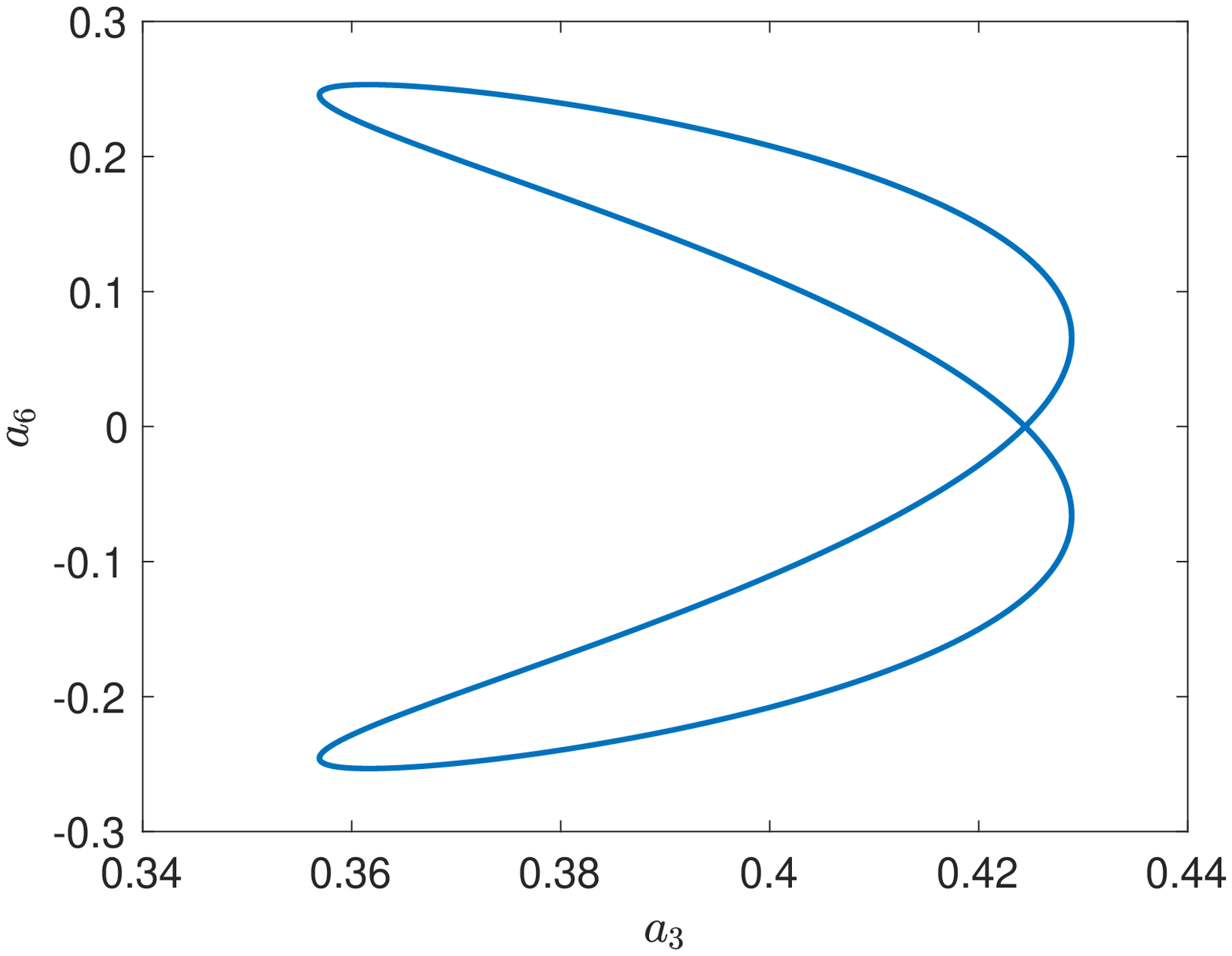}\caption{Phase portrait of modes 3 and 6}}\end{subfigure}
\centerline{\begin{subfigure}{0.5\textwidth}{\includegraphics[width=1.0\textwidth]{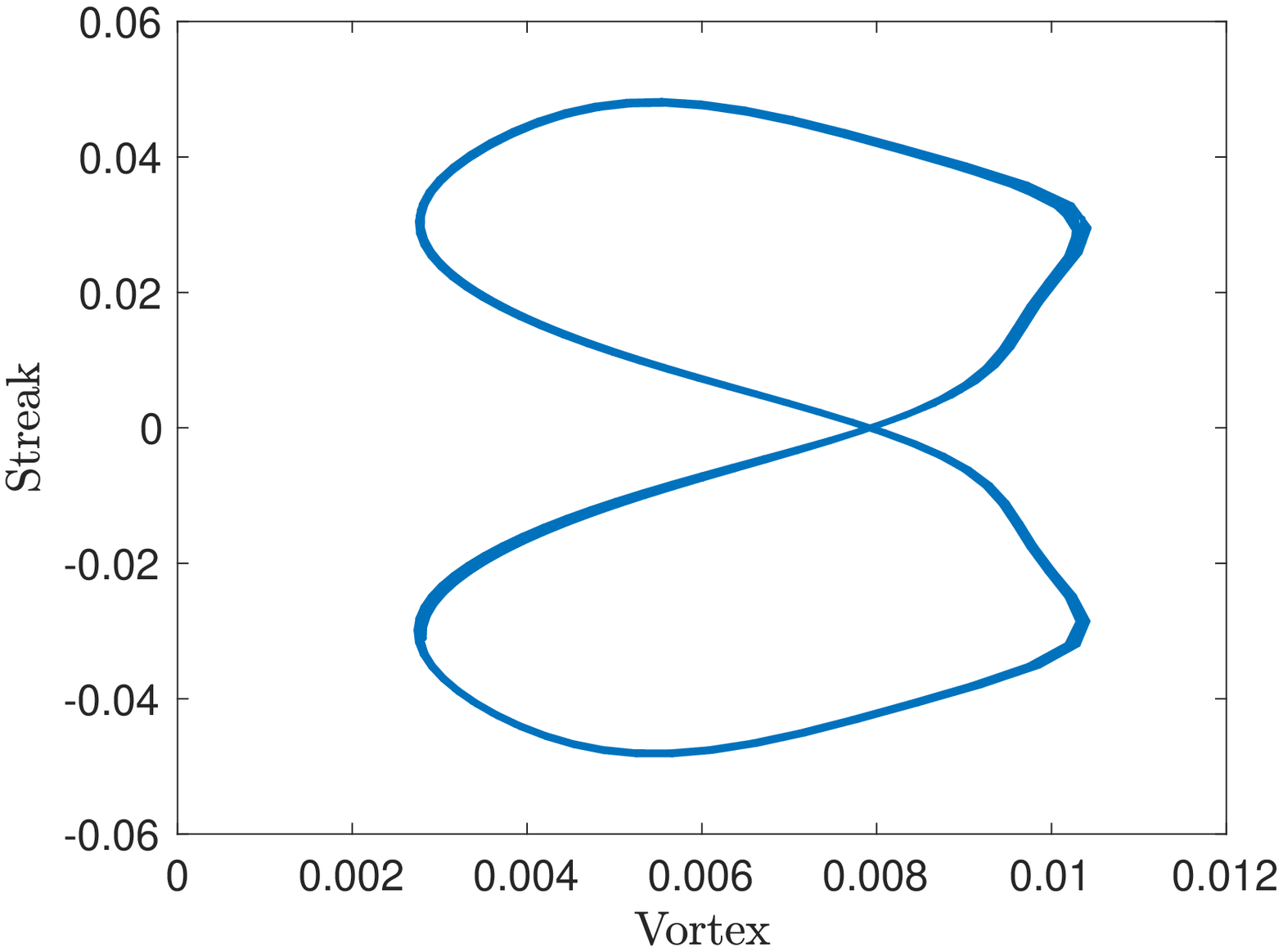}\caption{DNS phase portrait of vortices and streaks, from NC}}\end{subfigure}}
\caption{Limit-cycle oscillation for $Re=50$ and comparison with DNS results from NC ($Re=200$ and non-slip boundary conditions).}
\label{fig:attractors_first_LCO_timeseries}
\end{figure}

The behaviour of the system as a function of time for $\mathrm{Re}=50$, after the Hopf bifurcation, is illustrated in figure \ref{fig:attractors_first_LCO_timeseries}. The system has a similar behaviour of what was found in NC. Modes $a_1$ (mean-flow 1) and $a_6$ (streak) oscillate, with low values of $a_1$ occurring for large $|a_6|$, which can be thought of high-amplitude streaks and rolls (modes $a_5$ and $a_6$) leading to strong mean-flow distortions, reducing the mean shear. In turn, low (high) mean shear stabilises (destabilises) the vortex modes $a_3$ and $a_4$, which have an amplitude decay (growth). This has led the labelling of this cyclic behaviour as permanently actuated, periodicaly unstable (PAPU) flow in NC, as the observation of the KH modes $a_3$ and $a_4$ alone gives the impression that the system oscillates between KH-stable and KH-unstable. In dynamical systems theory, the observed behaviour is a stable limit cycle.

A phase portrait involving $a_3$ and $a_6$ is also shown in figure \ref{fig:attractors_first_LCO_timeseries}(b). The portrait has a distorted ``8'' shape, with streaks $a_6$ changing in sign and an amplitude variation of KH mode $a_3$ without sign change. {For comparison, a phase portrait taken from the DNS in NC is shown in figure \ref{fig:attractors_first_LCO_timeseries}(c), with "streak" and "vortex" states obtained using the streamwise velocity at wavenumber $(k_x,k_z)=(0,\gamma)$ and $(\alpha,0)$, respectively, at $y=0.5$. As the DNS in NC uses non-slip boundary conditions, and is carried out at $Re=200$, only a qualitative comparison is possible, and thus the velocity fluctuations at this position and wavenumbers were deemed sufficient to evaluate a corresponding phase portrait from the DNS data.} For both the present ROM and the NC data, large vortex amplitudes $a_3$ occur for low values of streak amplitude $|a_6|$, in the middle section of the ``8''.   Analysis of streaky shear layers~\citep{marant2018influence} and jets~\citep{lajus2019spatial,wang2021effect}, with steady streaks considered in a base flow, shows that they have a stabilising effect on the Kelvin-Helmholtz mechanism. The observed limit cycle, as the one in NC, displays a similar behaviour, albeit in the non-linear dynamics.

\begin{figure}
\begin{subfigure}{0.5\textwidth}{\includegraphics[width=1.0\textwidth]{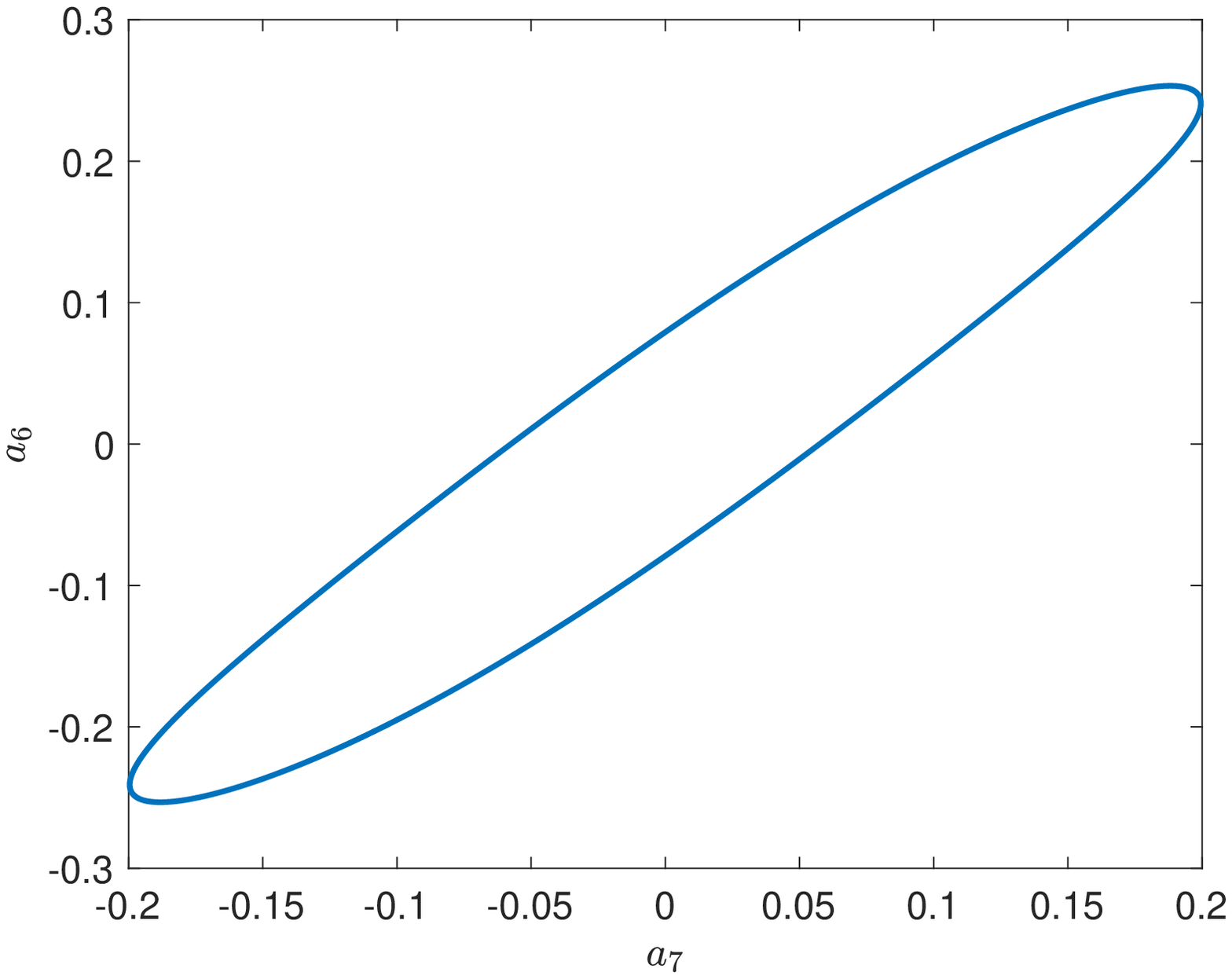}\caption{Phase portrait of modes 7 and 6}}\end{subfigure}
\begin{subfigure}{0.5\textwidth}{\includegraphics[width=1.0\textwidth]{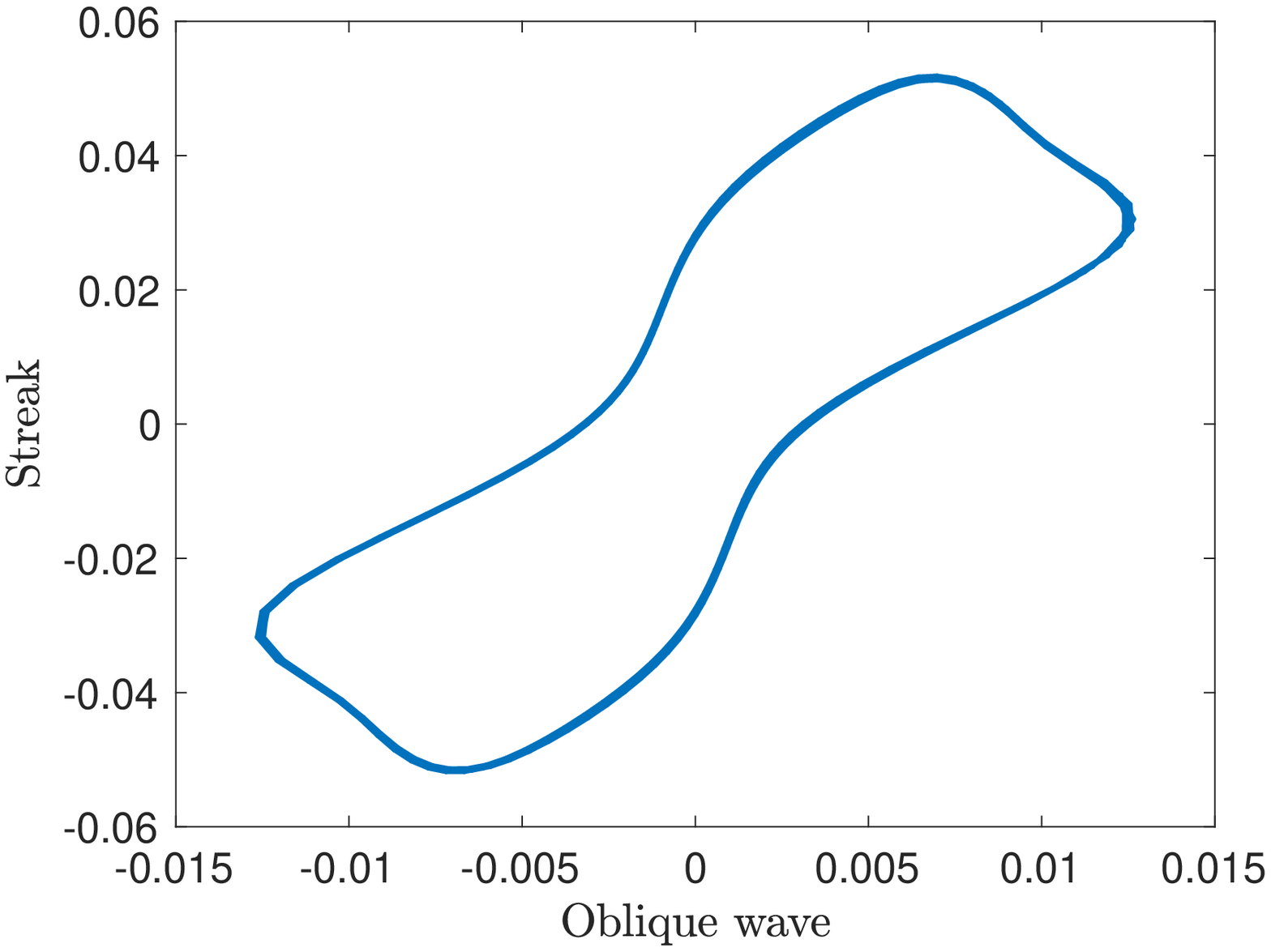}\caption{DNS phase portrait of streaks and oblique waves}}\end{subfigure}
\caption{Phase portraits of streaks and oblique waves in the limit-cycle oscillation for $Re=50$ in the ROM, and $Re=200$ for the NC DNS data.}
\label{fig:attractors_first_LCO_streakoblique}
\end{figure}

Oblique waves have an important role in defining the periodic behaviour of the model. As explored in NC, these waves grow exponentially with the vortices, and their phase is closely associated to the phase of streaks, such that alternation between positive and negative streaks is directly associated to the phase of these waves. This is exemplified in figure \ref{fig:attractors_first_LCO_streakoblique}, {displaying phase portraits from the model and from the DNS of NC. In the latter case, the "oblique wave" state in the DNS was obtained by taking the streamwise velocity with $(k_x,k_z) = (\alpha,\gamma)$  at $y=-0.5$. As in the results of figure \ref{fig:attractors_first_LCO_timeseries}, the comparison is only qualitative due to the different Reynolds number and boundary conditions, but the phase portraits show similar patterns, with a clear correlation between streaks and oblique waves.}

The significance of this limit cycle for the dynamics and sound radiation of jets is that it displays amplitude modulation of the KH mode $a_3$ in time. This is known to be a mechanism of sound generation \citep{sandham2006sound}, especially when occurring simultaneously with amplitude modulation in space \citep{cavalieri2011jittering}. In a shear layer or jet, the latter is known to occur due to the spatial spreading of the shear layer, leading to the spatial decay of the Kelvin-Helmholtz mode in downstream regions~\citep{jordan2013wave,cavalieri2019wave}. Due to the homogeneity in $x$ of the present confined shear layer, represented by modes that are strictly periodic in $x$, an amplitude modulation in space is not possible. However, the observation of temporal modulation is relevant, and may be related to observations in turbulent jets, as discussed in NC.


Notice finally that, due to the symmetry of the system, there is another stable limit cycle with $a_3<0$, which bifurcates from the lower branch of the vortex solutions in figure \ref{fig:vortex_pitchfork}.

\subsection{Transition to chaos for $P=0.08$}
\label{sec:transitionchaos}

\begin{figure}
\begin{subfigure}{0.5\textwidth}{\includegraphics[width=1.0\textwidth]{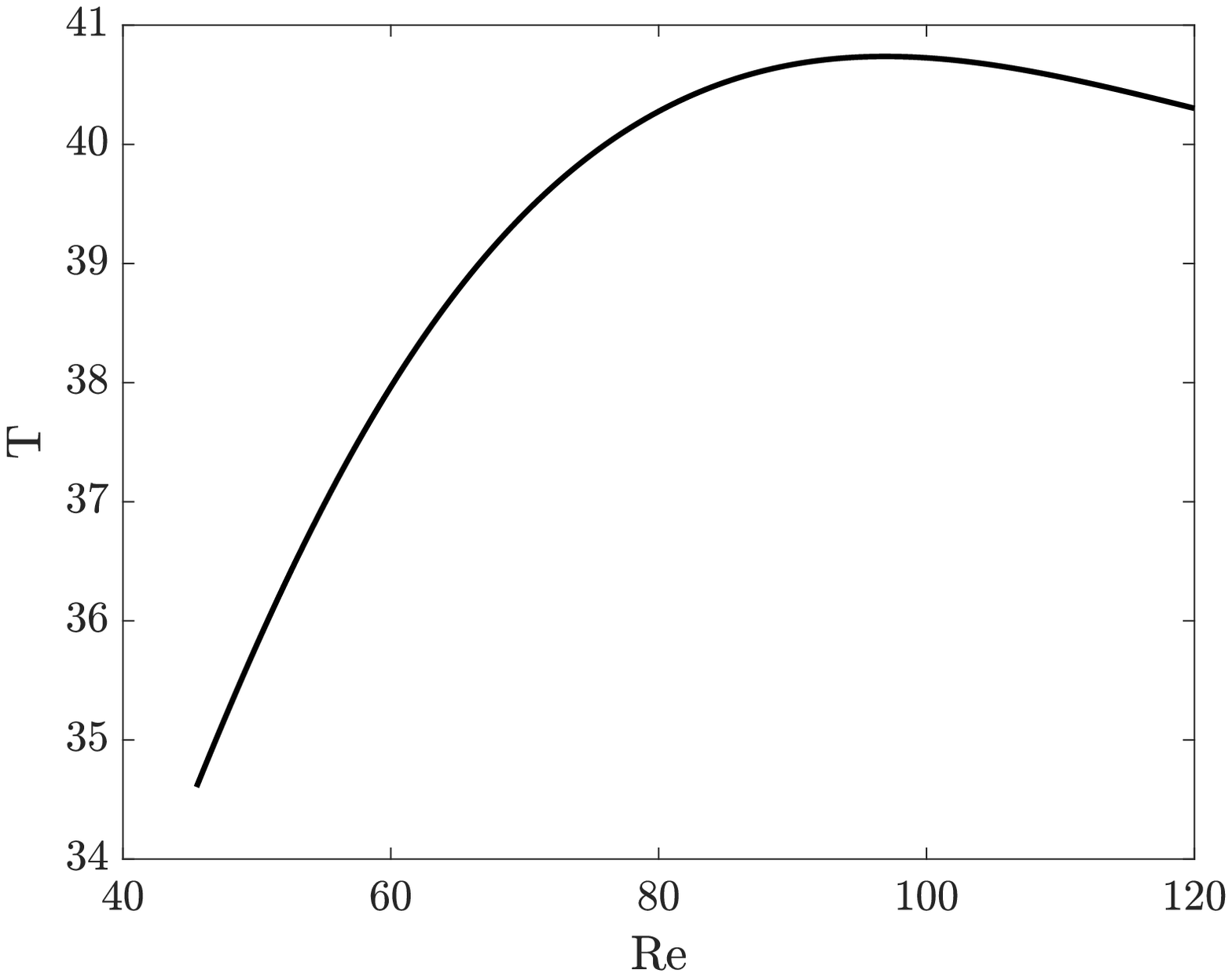}}\caption{Period of the cycle}\end{subfigure}
\begin{subfigure}{0.5\textwidth}{\includegraphics[width=1.0\textwidth]{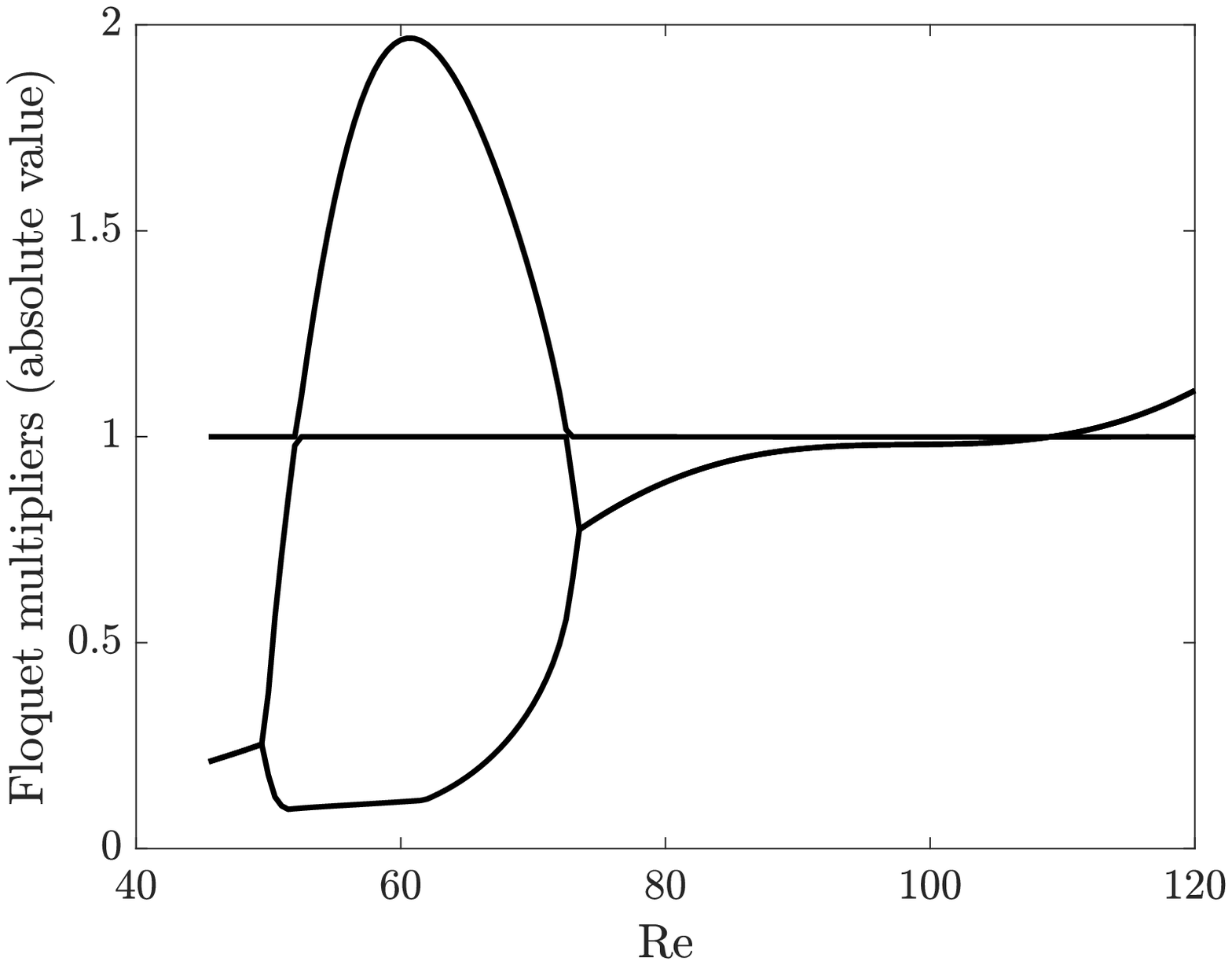}\caption{Abs. value of 3 leading Floquet multipliers}}\end{subfigure}
\caption{Characteristics of the cycle emerging from the Hopf bifurcation at $\mathrm{Re}=41.8$.}
\label{fig:floquet}
\end{figure}

As Re is increased beyond 41.8, the limit cycle of figure \ref{fig:attractors_first_LCO_timeseries} develops with increasing amplitude. This cycle was tracked until $\mathrm{Re} = 120$ by looking for time-periodic solutions of eq. (\ref{eq:ROM}). The cycle stability was analysed using Floquet theory, as described by \cite{kawahara2012significance}. The period of the limit cycle and its leading Floquet multipliers are shown in figure \ref{fig:floquet}. There is always one multiplier equal to 1, corresponding to the neutral direction following the periodic orbit. With increasing $\mathrm{Re}$ we observe a  first instability appearing between $\mathrm{Re}=52$ and 72.5. Inspection of the unstable Floquet multiplier shows that it is real and positive, and we thus have a crossing of the unit circle at $+1$. The cycle stability is restored at $\mathrm{Re}=72.5$, and a secondary Hopf bifurcation, with complex conjugate Floquet multipliers leaving the unit circle, happens at $Re=109$. These instabilities are relevant for the appearance of chaotic behaviour, as will be shown in the next subsections.

\subsubsection{Symmetry breaking}
\label{sec:symbreak}

The numerical solutions of the system in this range of $\mathrm{Re}$ show that we have a symmetry breaking bifurcation at $\mathrm{Re}=52$, followed by symmetry restoring at $\mathrm{Re}=72.5$. To show this behaviour, a Poincar\'e section is defined at the hyperplane $a_5=0$, and we collect points for the non-linear simulation crossing the plane with $\dot{a_5}<0$. Thus the behaviour of the dynamical system may be tracked with the Poincar\'e section that defines a map. This is carried out successively for Reynolds number varying in small increments, with the final state of the previous simulation taken as initial condition for the following one to minimise transients. Simulations with duration $T=2000$ are carried out and the points in the Poincar\'e section are kept only for the final half of the simulation. With this procedure we analyse systematically the attractor for each $\mathrm{Re}$, without significant transient effects. The resulting points in the Poincar\'e section can be plotted as a function of Reynolds number, illustrating the changes in the behaviour of the system: a stable limit cycle becomes a fixed point in the Poincar\'e section; in case of period doubling, a period-$n$ periodic orbit appears as $n$ points in the section; a quasi-periodic attractor is a dense region; and chaos is also manifest by the lack of periodicity. The same approach was used by \cite{kashinath2014nonlinear} in the analysis of the nonlinear dynamics of a thermoacoustic system. As discussed previously, the present system has two limit cycles, one with $a_3>0$ and another with $a_3<0$. The following results show only the dynamics related to the $a_3>0$; however, mirror solutions with $a_3<0$ also exist due to the symmetry of the problem.

\begin{figure}
\begin{subfigure}{0.5\textwidth}{\includegraphics[width=1.0\textwidth]{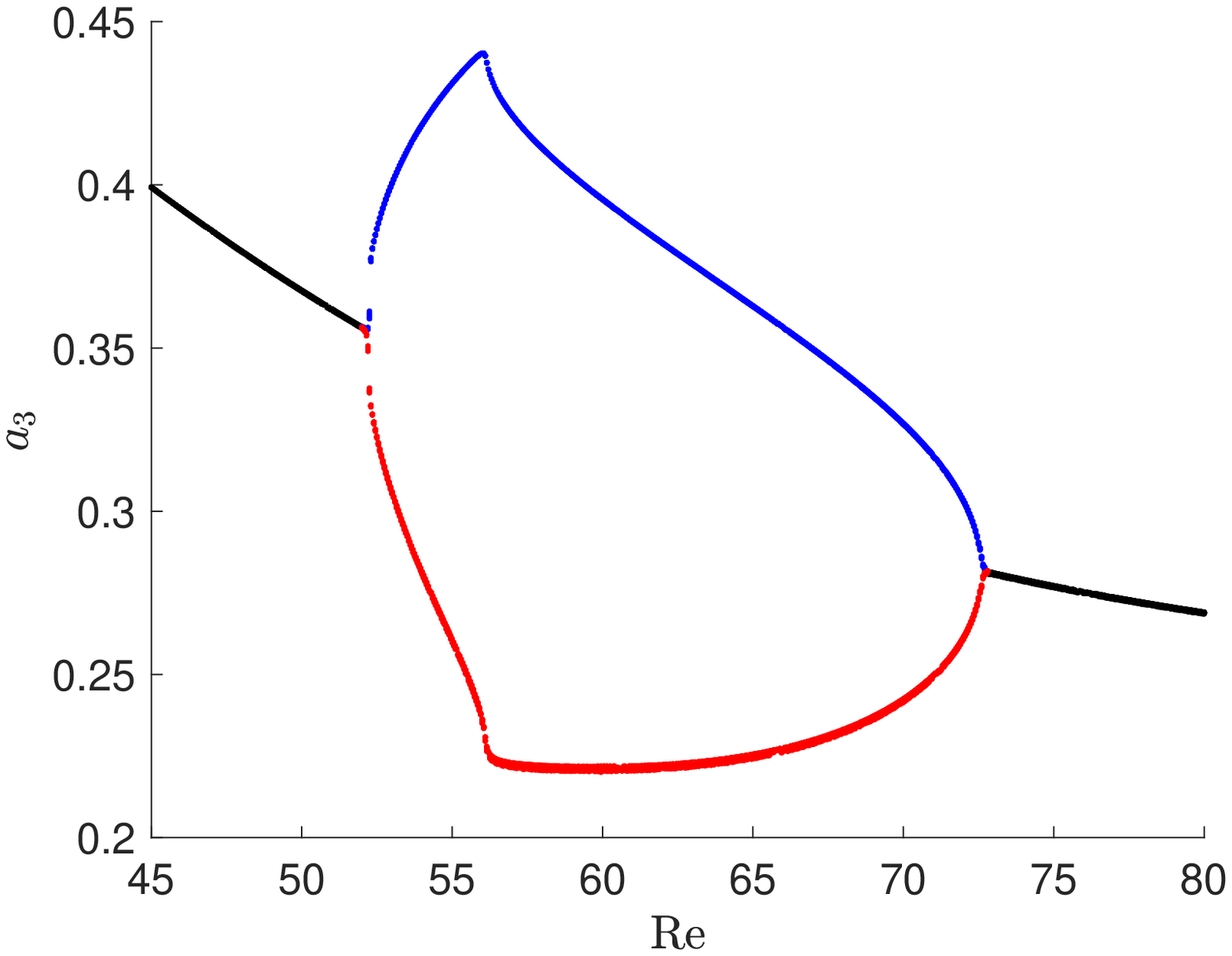}\caption{Poincar\'e section}}\end{subfigure}
\begin{subfigure}{0.5\textwidth}{\includegraphics[width=1.0\textwidth]{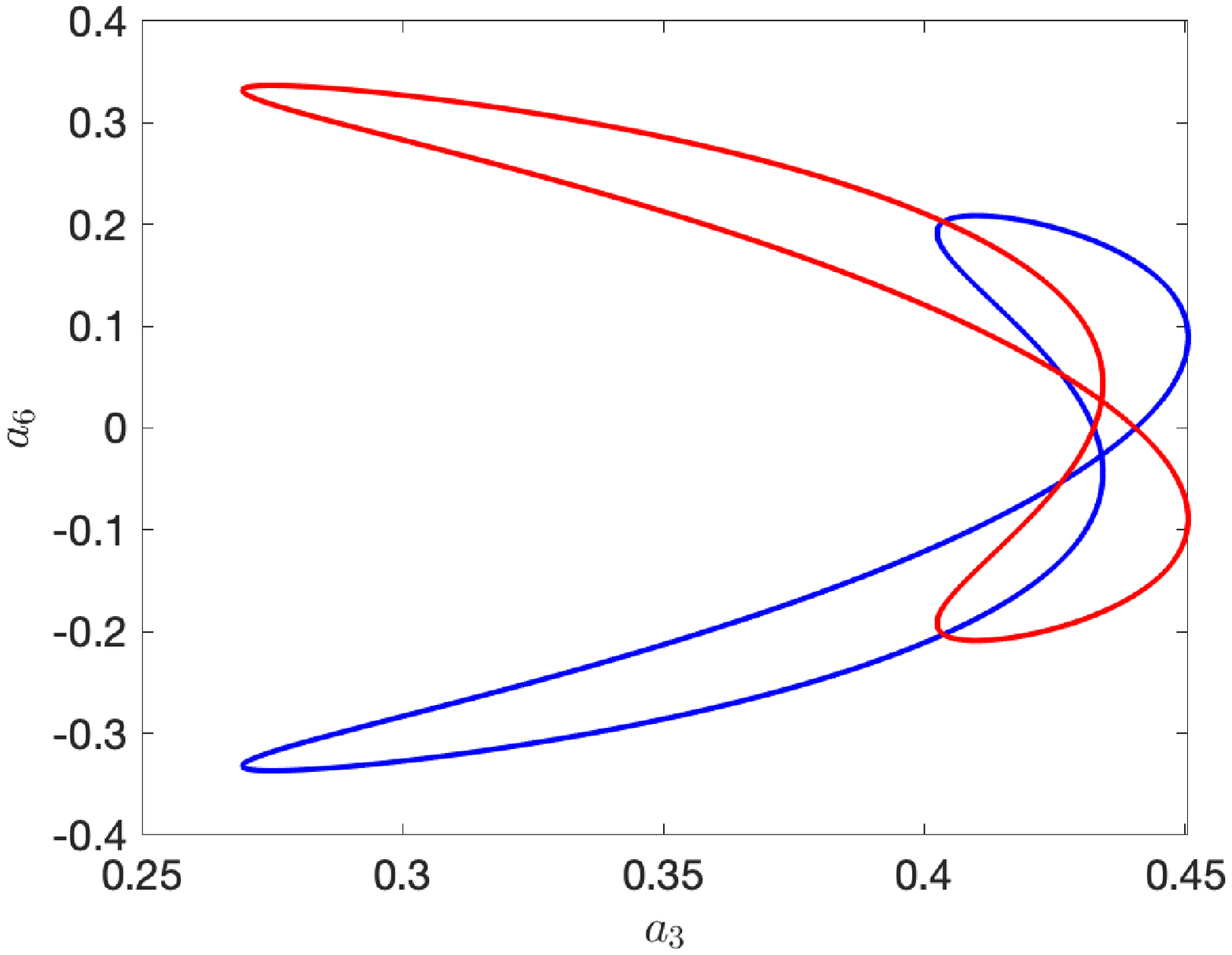}\caption{Phase portrait for $\mathrm{Re}=54$}}\end{subfigure}
\caption{Illustration of the symmetry breaking and restoring bifurcations. Phase portraits in (b) correpond to the red and blue branches shown in (a).}
\label{fig:symbreak}
\end{figure}

The Poincar\'e section for $\mathrm{Re}$ between 42 and 80 is shown in figure \ref{fig:symbreak}(a). For $\mathrm{Re}<52$ the system has a single stable limit cycle, which becomes a point in the Poincar\'e section for each $\mathrm{Re}$.  As the Reynolds number is increased beyond 52, the cycle becomes unstable and two other periodic solutions emerge in a symmetry-breaking bifurcation. A sample phase portrait for $\mathrm{Re}=54$, is shown in figure \ref{fig:symbreak}, where we see that the symmetry of the ``8'' shape is broken. Such asymmetric solutions suffer a symmetry-restoring bifurcation at  $\mathrm{Re=72.8}$, and the limit cycle analysed in \ref{sec:bifurcations} becomes stable again. Both bifurcations are supercritical, as they do not show hysteresis. 

\subsubsection{Formation of a chaotic saddle}
\label{sec:chaoticsaddle}

\begin{figure}
\begin{subfigure}{0.5\textwidth}{\includegraphics[width=1.0\textwidth]{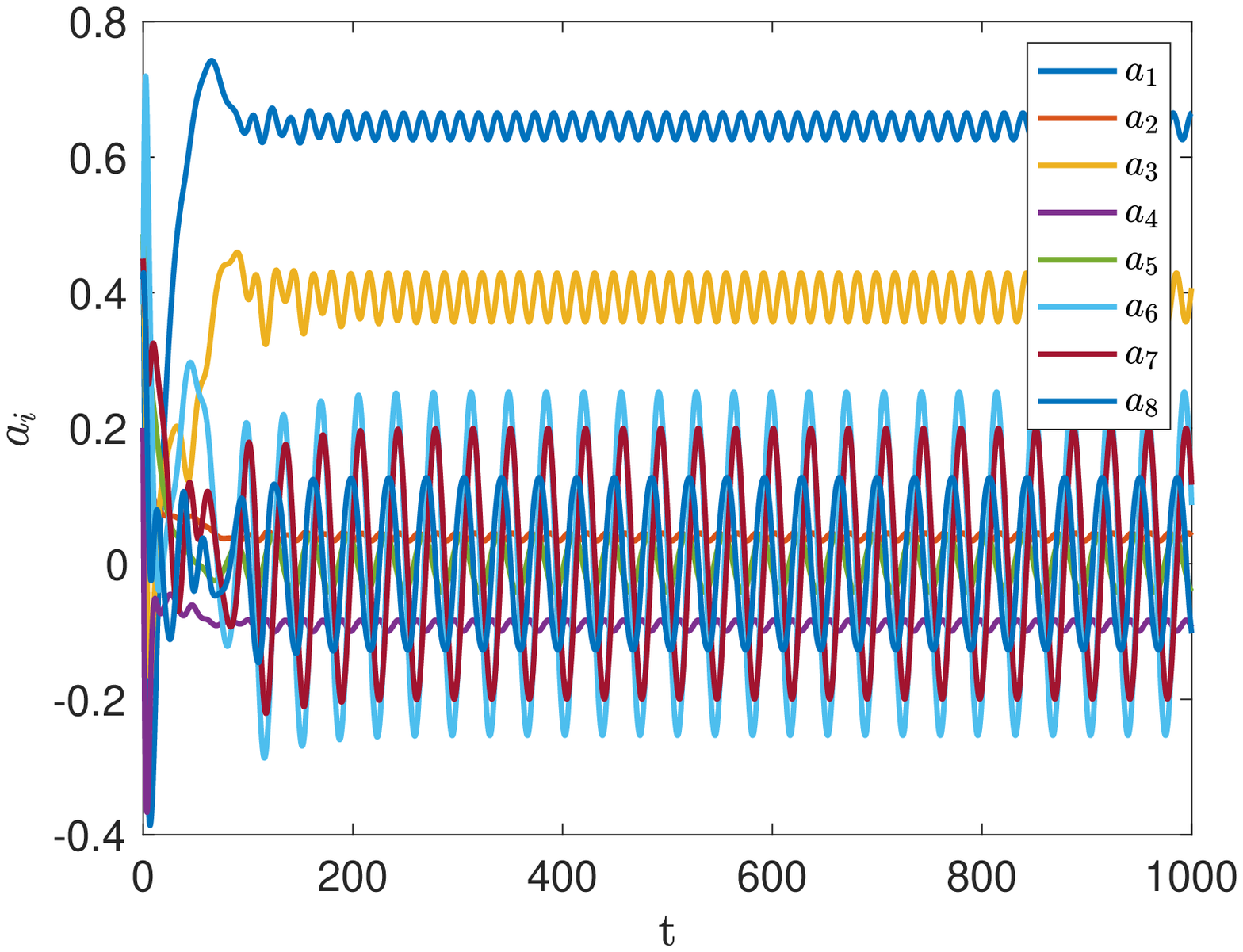}}\caption{$\mathrm{Re}=50$}\end{subfigure}
\begin{subfigure}{0.5\textwidth}{\includegraphics[width=1.0\textwidth]{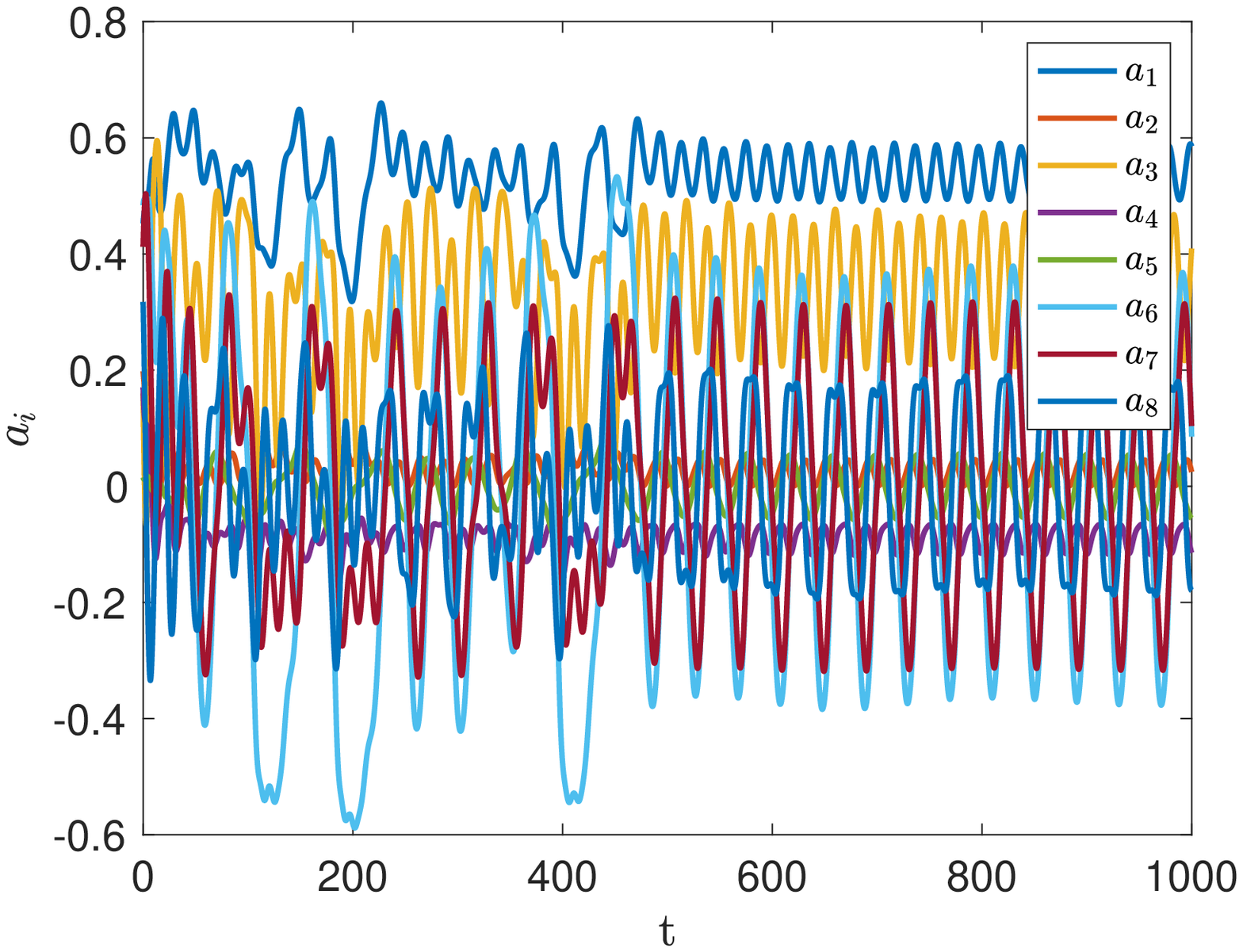}}\caption{$\mathrm{Re}=80$}\end{subfigure}
\caption{Transients leading to time-periodic behaviour. The system displays transient chaos for $\mathrm{Re}=80$.}
\label{fig:transientchaos}
\end{figure}

A difference between the dynamics before and after symmetry breaking and restoring is the appearance of transient chaos. This behaviour is illustrated in figure \ref{fig:transientchaos}. Whereas the system quickly settles into the limit cycle for $\mathrm{Re}=50$, for $\mathrm{Re}=80$ the modes display chaotic oscillations for a long transient before finally settling into the stable limit cycle. This suggests the existence of a chaotic saddle, similar to observations in reduced-order models of wall turbulence \citep{eckhardt1999transition,moehlis2004low}. Such saddle attracts neighbouring states by its stable manifold, and the dynamics remain in the close vicinity of the saddle for a long transient until it is eventually repelled by the unstable manifold. 

\begin{figure}
\begin{subfigure}{0.5\textwidth}{\includegraphics[width=1.0\textwidth]{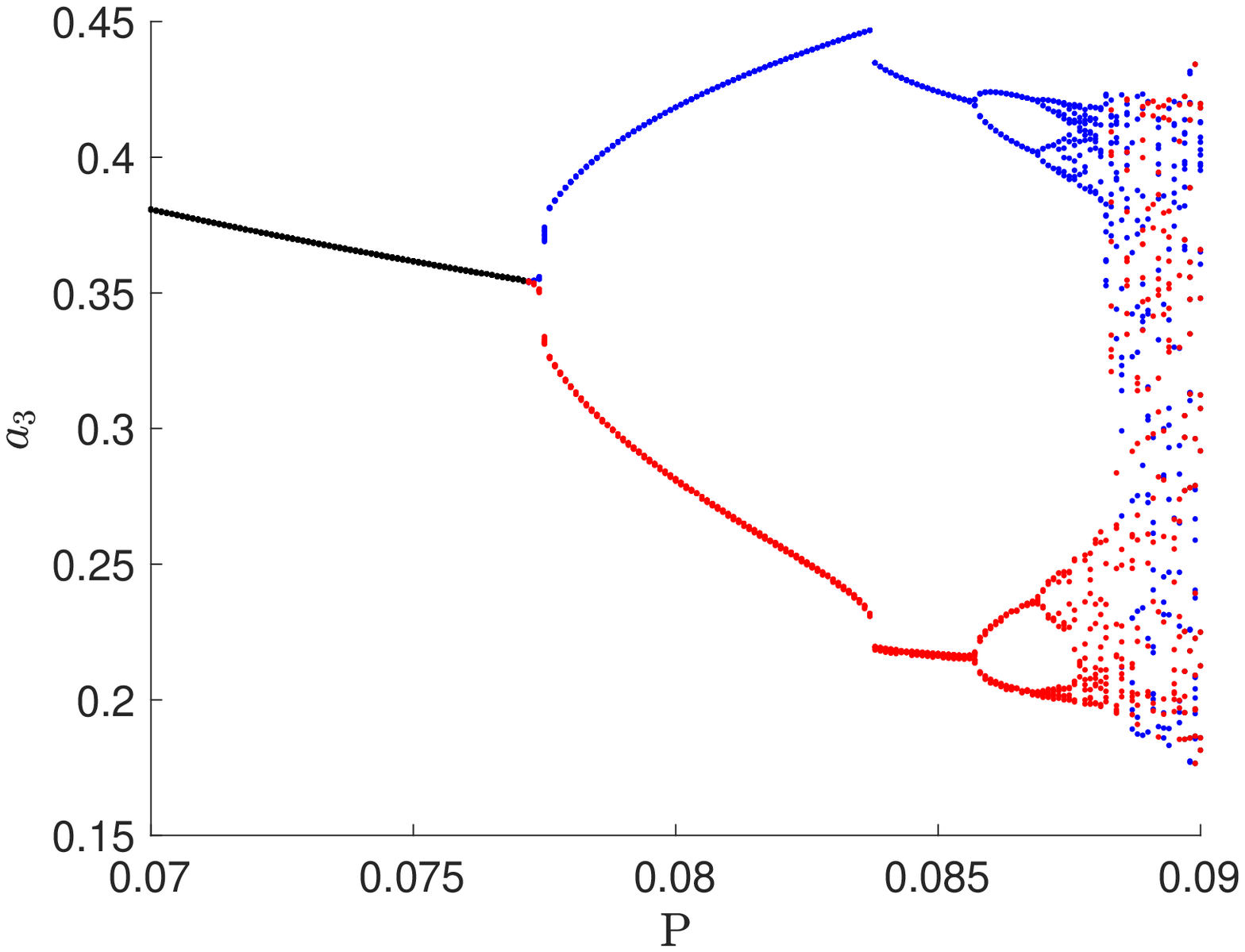}\caption{Full view}}\end{subfigure}
\begin{subfigure}{0.5\textwidth}{\includegraphics[width=1.0\textwidth]{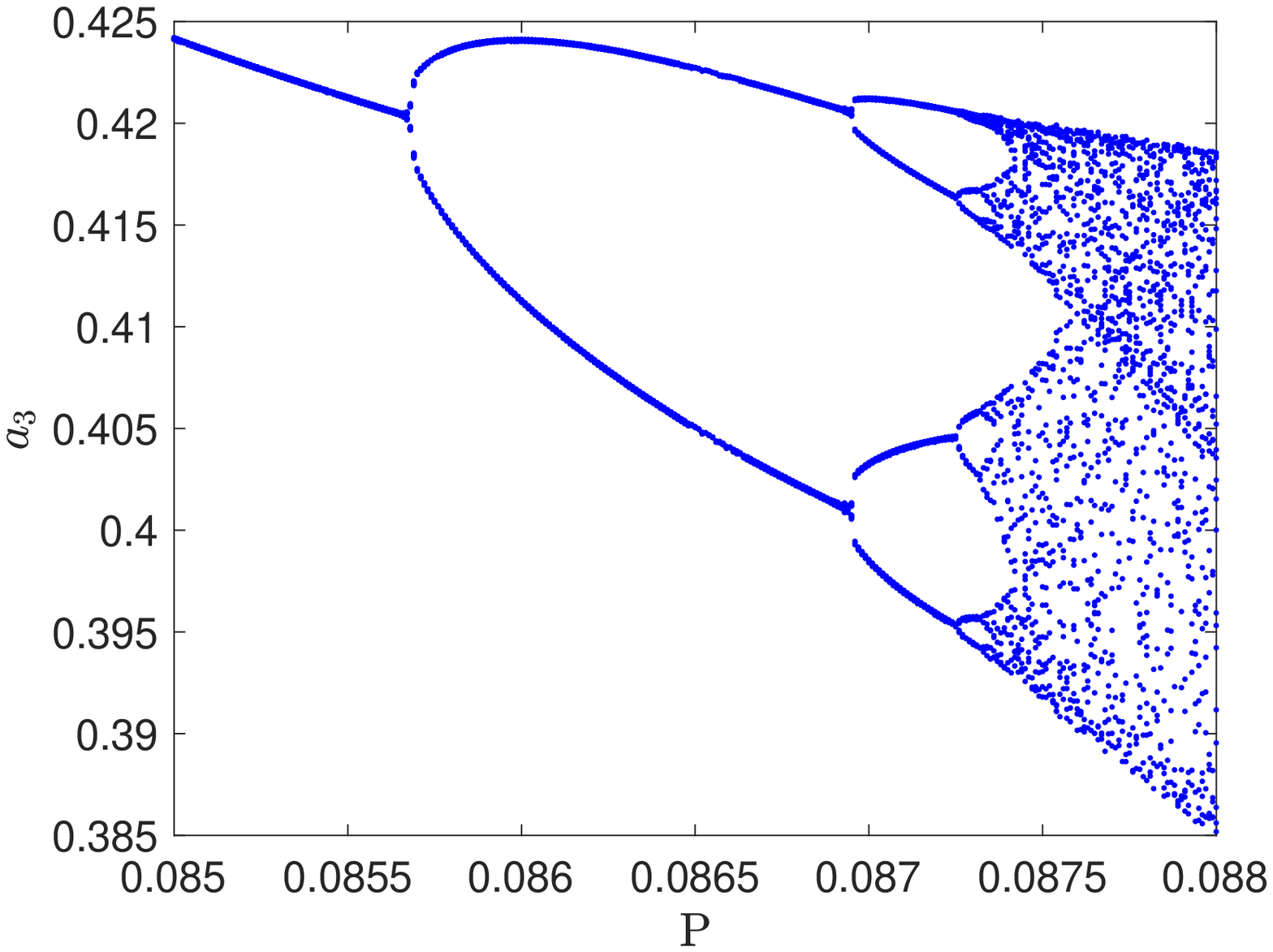}\caption{Detail of period-doubling cascade}}\end{subfigure}
\caption{Poincar\'e section for $\mathrm{Re}=54$ and varying $P$. Blue and red dots show sections tracked with one of the two asymmetric limit cycles of figure \ref{fig:symbreak}.}
\label{fig:creation_saddle}
\end{figure}

The origin of the chaotic saddle may be tracked by following a Poincar\'e section with fixed $\mathrm{Re}$ and varying $P$. This was carried out for $\mathrm{Re}=54$, which for $P=0.08$ has two asymmetric limit cycles, as shown in \ref{sec:symbreak}. The Poincar\'e section, calculated as in figure \ref{fig:symbreak} is shown in figure \ref{fig:creation_saddle}(a). As $P$ is increased with fixed $\mathrm{Re}$, the system undergoes a symmetry-breaking bifurcation for $P=0.077$. A further increase of $P$ leads to a saddle-node bifurcation for $P=0.0837$, seen in figure \ref{fig:creation_saddle}(a) as a jump in the value of $a_3$ in the Poincar\'e section. This bifurcation is not primarily relevant for the emergence of a chaotic saddle; further details are shown in the Appendix, where it is shown that the jump in $a_3$ is related to two saddle-node bifurcations. The limit cycles for $P>0.0837$ remain asymmetric as in figure \ref{fig:symbreak}. For $P=0.0857$ the cycles undergo a period-doubling bifurcation, starting Feigenbaum cascades leading to chaos for $P\approx 0.0874$. The cascade for one of the attractors is shown in more detail in figure \ref{fig:creation_saddle}(b).

\begin{figure}
\begin{subfigure}{0.5\textwidth}\includegraphics[width=1.0\textwidth]{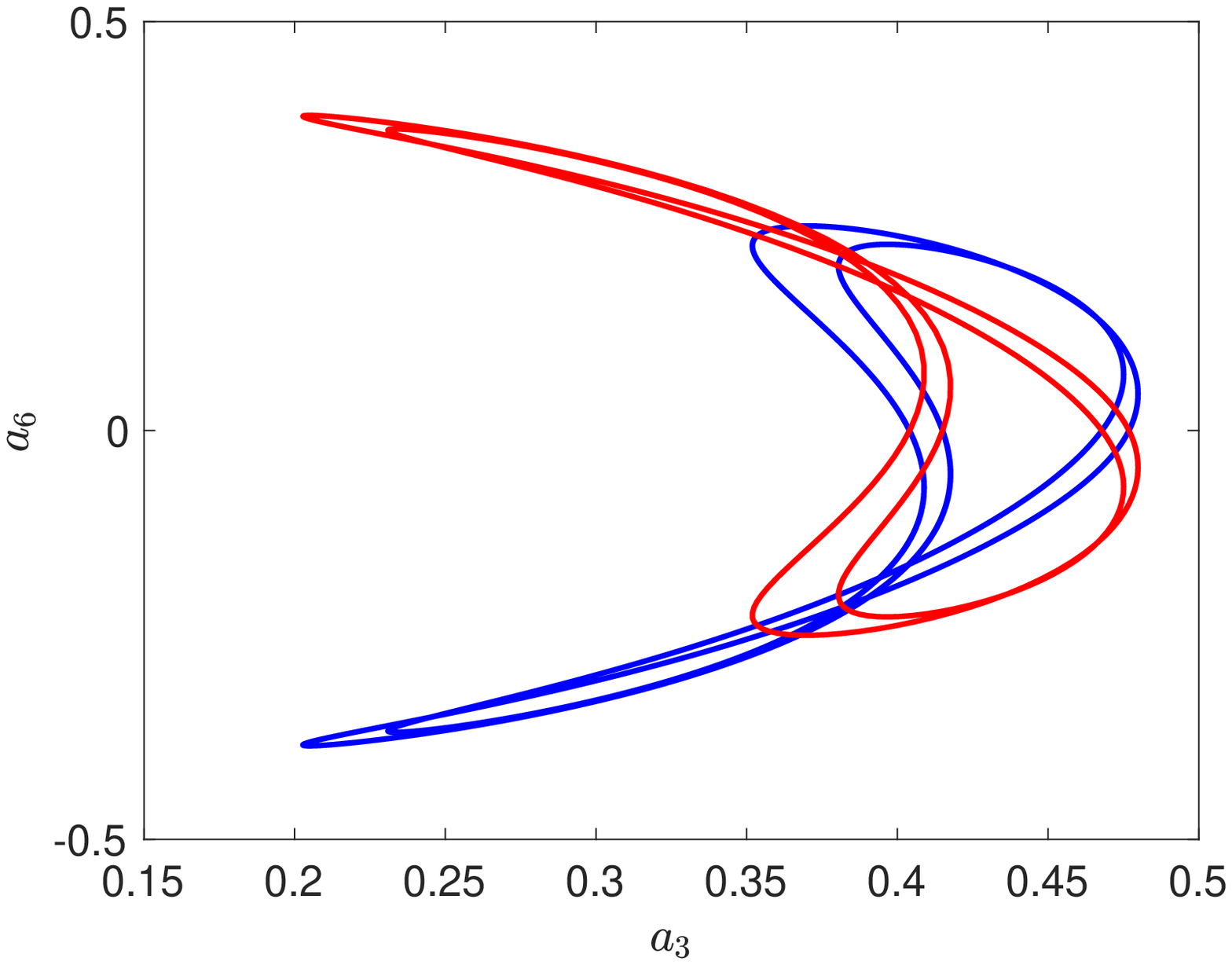}\caption{$P=0.0865$, period-2 cycle}\end{subfigure}\begin{subfigure}{0.5\textwidth}\includegraphics[width=1.0\textwidth]{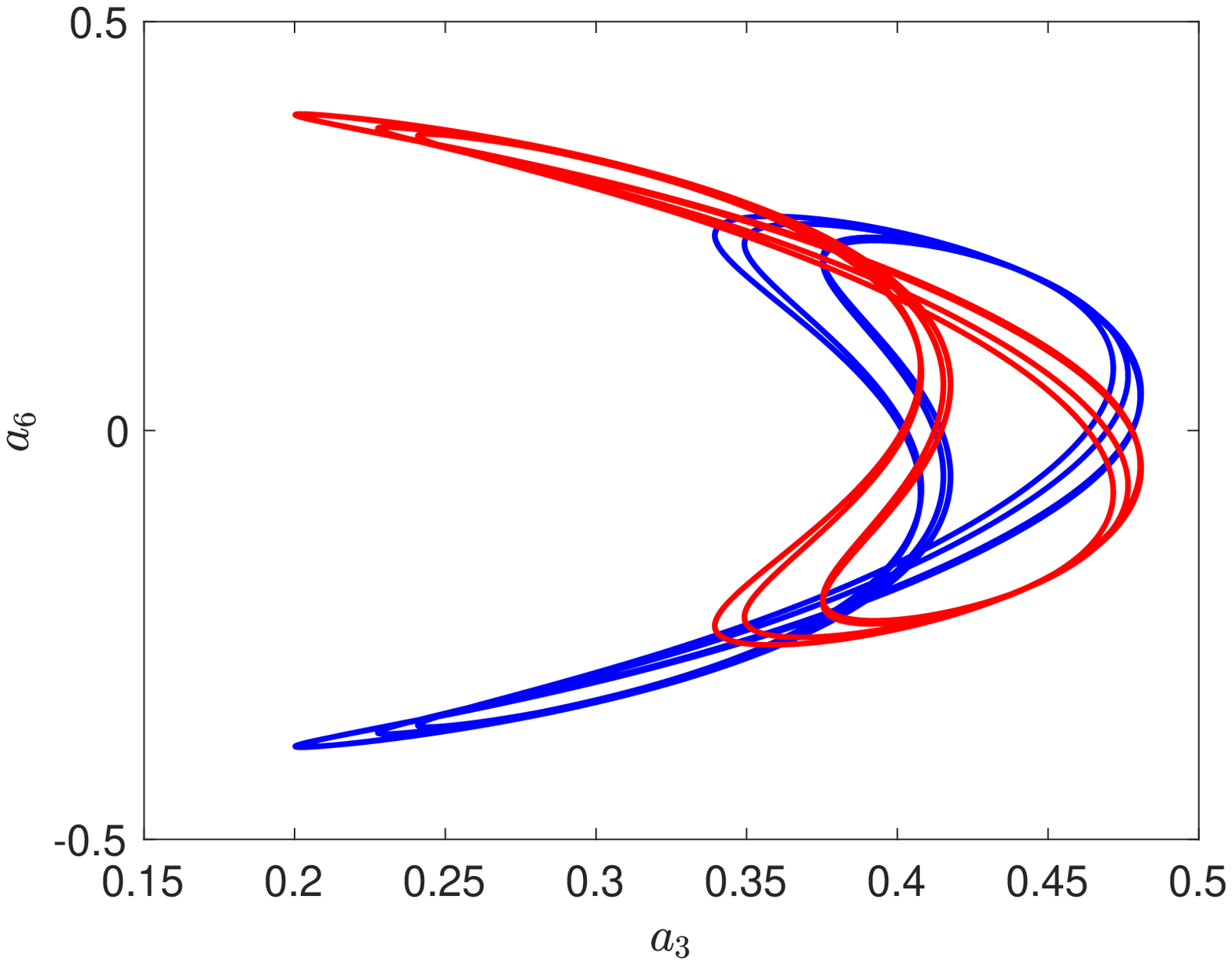}\caption{$P=0.0872$, period-4 cycle}\end{subfigure}
\begin{subfigure}{0.5\textwidth}\includegraphics[width=1.0\textwidth]{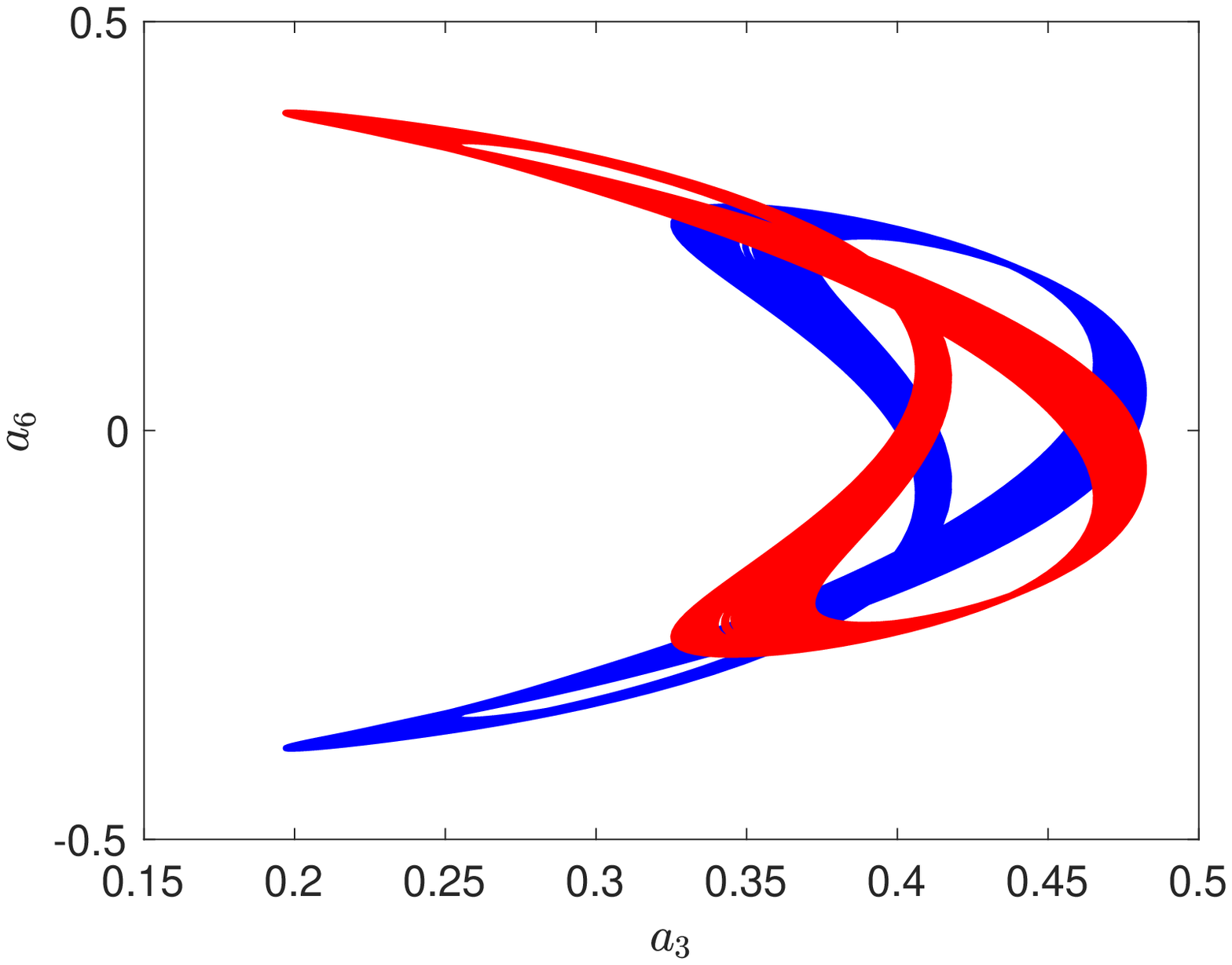}\caption{$P=0.088$, chaotic attractors}\end{subfigure}\begin{subfigure}{0.5\textwidth}\includegraphics[width=1.0\textwidth]{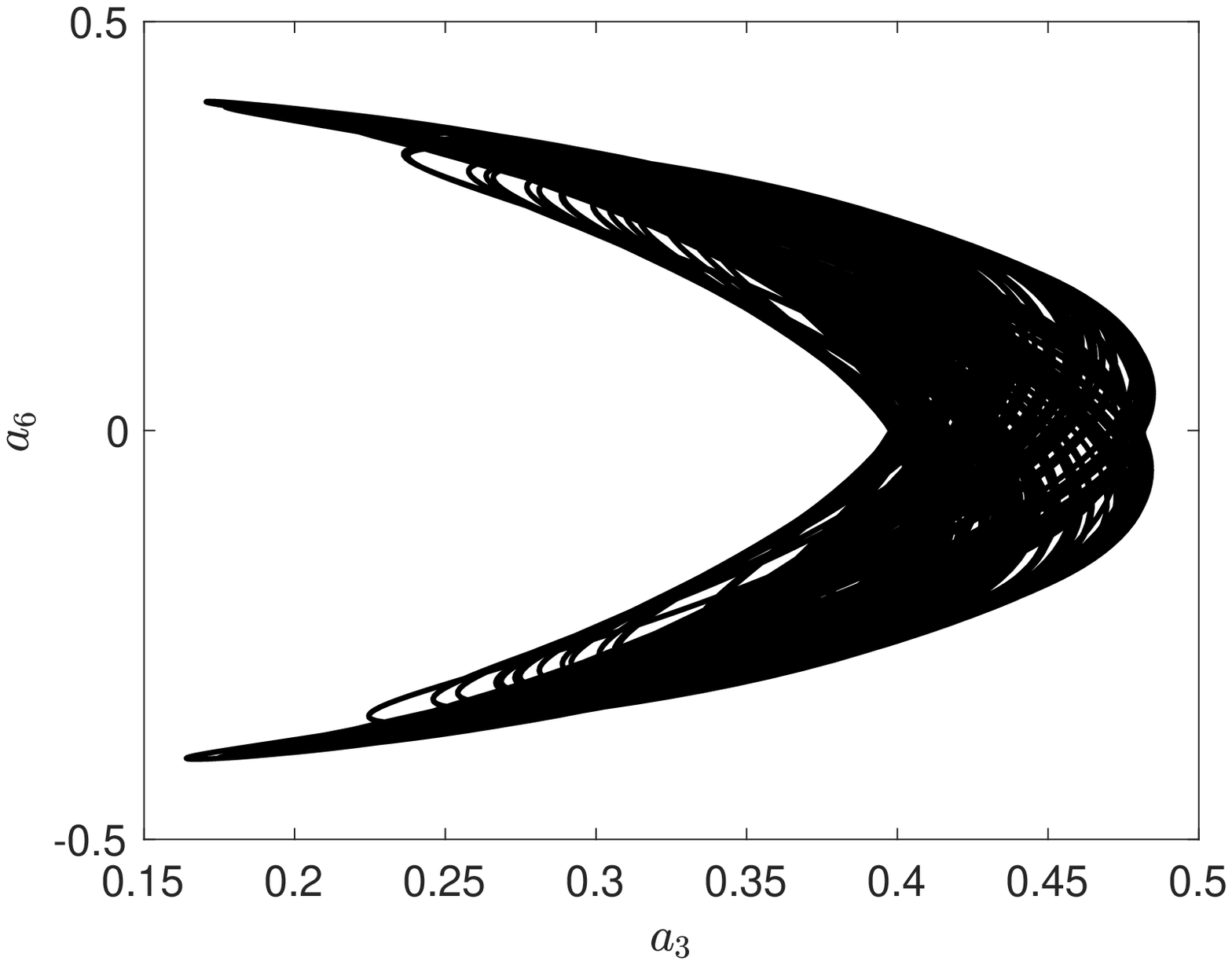}\caption{$P=0.09$, merged chaotic attractor}\end{subfigure}
\caption{Phase portraits of attractors for $\mathrm{Re}=54$ and selected $P$. Blue and red lines show attractors with the convention of figure \ref{fig:creation_saddle}. The larger attractor for $P=0.09$ results from the merging crisis of the individual chaotic attractors shown in (c).}
\label{fig:chaotic_saddle_attractors}
\end{figure}

Up to $P=0.088$ the two attractors remain distinct, with chaos confined to a low range of values of $a_3$. For $P$ beyond this value, the chaotic attractors undergo a merging crisis, as they simultaneously collide with their basin boundaries. This is seen in figure \ref{fig:creation_saddle}(a) as it becomes no longer possible to distinguish between blue and red dots. The merging crisis leads to an attractor that is larger than the previous ones combined, as observed by \cite{chian2005attractor}. Sample phase portraits are shown in figure \ref{fig:chaotic_saddle_attractors}, highlighting the sequence of events: the limit cycles undergo period doubling, as illustrated in figures \ref{fig:chaotic_saddle_attractors}(a) and (b), and become chaotic in (c). Such chaotic attractors remain distinct, and the system does not go from one to the other. A further increase of $P$ leads to the merging crises, with a larger merged attractor  in figure \ref{fig:chaotic_saddle_attractors}(d).

{We conjecture that the boundary between the basins of attraction of the blue and red pre-crisis attractors is fractal. It is a known fact in dynamical systems that fractal basin boundaries can be formed by the stable manifold of a chaotic saddle \citep{battelino1988multiple,rempel2008alfven}. At the merging crisis, both attractors simultaneously collide with the chaotic saddle and its stable manifold at the basin boundary and the three sets (two attractors and a chaotic saddle) merge to form the post-crisis chaotic attractor. The presence of the chaotic saddle is the reason why the post-crisis attractor is larger than the union of the two pre-crisis attractors.}

\begin{figure}
\begin{subfigure}{0.5\textwidth}\includegraphics[width=1.0\textwidth]{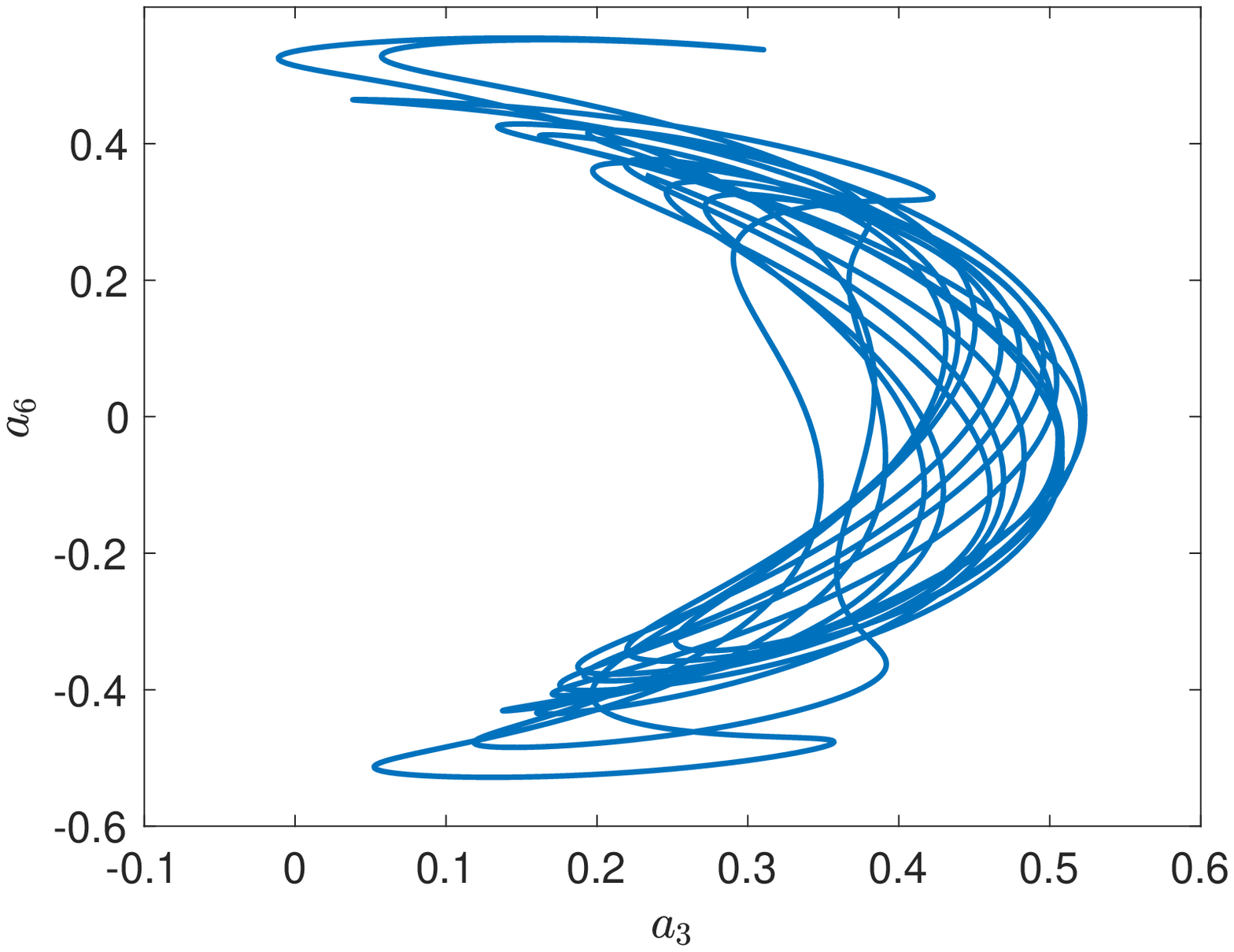}\caption{$P=0.08$, transient chaos}\end{subfigure}\begin{subfigure}{0.5\textwidth}\includegraphics[width=1.0\textwidth]{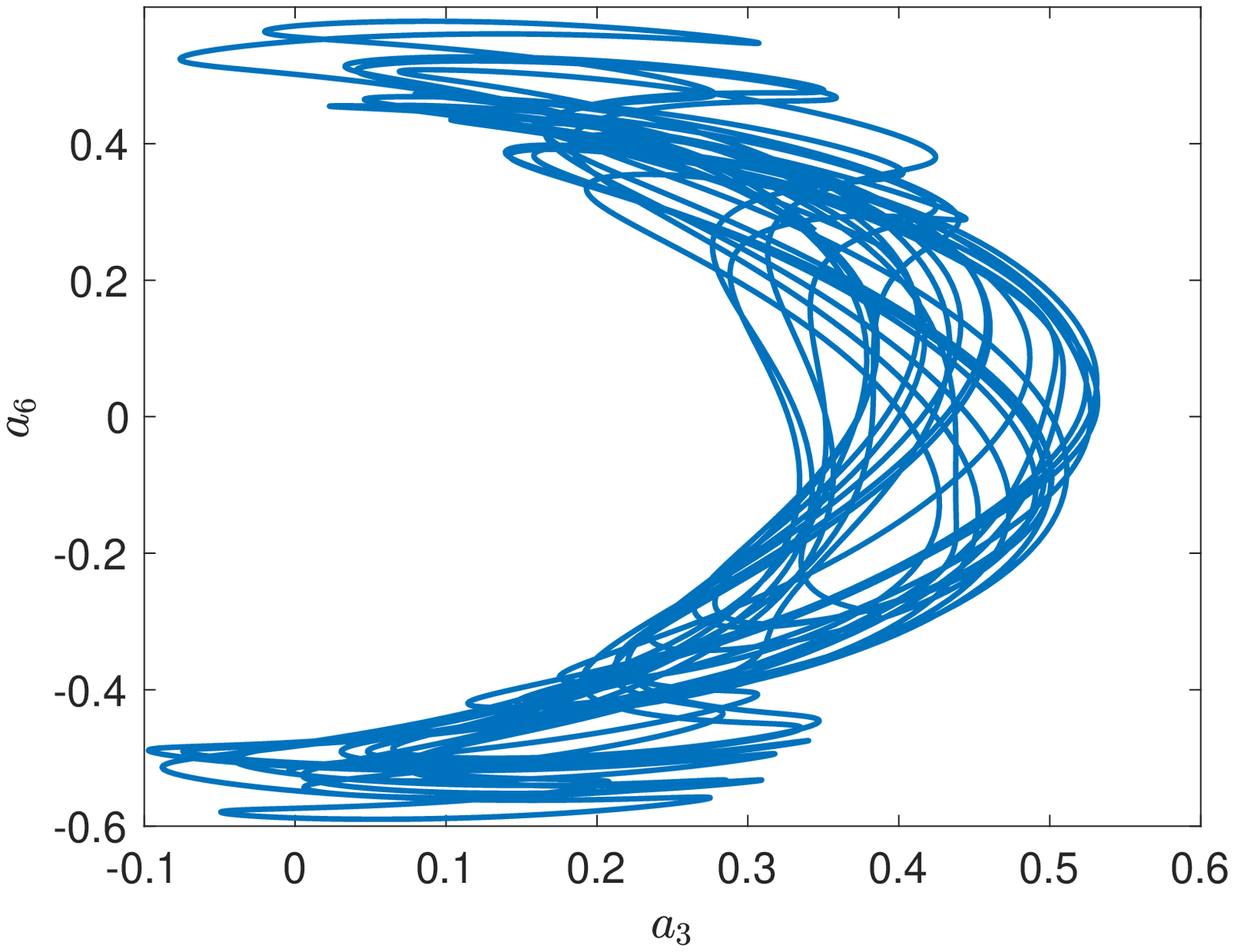}\caption{$P=0.09$, chaotic attractor}\end{subfigure}
\caption{Phase portraits for $Re=80$: (a) chaotic saddle with $P=0.08$ and (b) chaotic attractor with $P=0.09$.}
\label{fig:chaotic_saddle_Re80}
\end{figure}

{The period doubling cascade and subsequent merging crisis explain} the emergence of a chaotic attractor at larger $P$, for instance $P=0.09$ as shown in figure \ref{fig:creation_saddle} and \ref{fig:chaotic_saddle_attractors}(d). A chaotic attractor comprises an infinite number of unstable periodic orbits, and the dynamics are related to stable and unstable manifolds of such orbits~\citep{cvitanovic1989periodic}. For $P=0.08$ the dynamics eventually converge to the stable limit cycle; however, during transients, which become quite long for higher $\mathrm{Re}$, the behaviour is nonetheless chaotic, as illustrated in figure \ref{fig:transientchaos}(b). This may now be explained by the infinity of unstable periodic orbits that arise with the chaotic attractor for larger $P$. For $P=0.08$ such orbits are no longer able to retain indefinitely state-space trajectories, forming thus a chaotic saddle. However, during transients the system remains with dynamics similar to the ones observed for the chaotic attractor for $P=0.09$. This is illustrated in figure \ref{fig:chaotic_saddle_Re80}(a), which shows a phase portrait of a long transient taken for $P=0.08$ and $\mathrm{Re}=80$. The system has a trajectory that closely resembles the chaotic attractor for $P=0.09$, shown in figure \ref{fig:chaotic_saddle_Re80}(b). Hence, the high-P chaotic attractor {loses stability and becomes a saddle} for $P=0.08$, leading to the observed transients.

\subsubsection{Quasi-periodic and chaotic attractors}
\label{sec:finalchaos}

\begin{figure}
\begin{subfigure}{0.5\textwidth}\includegraphics[width=1.0\textwidth]{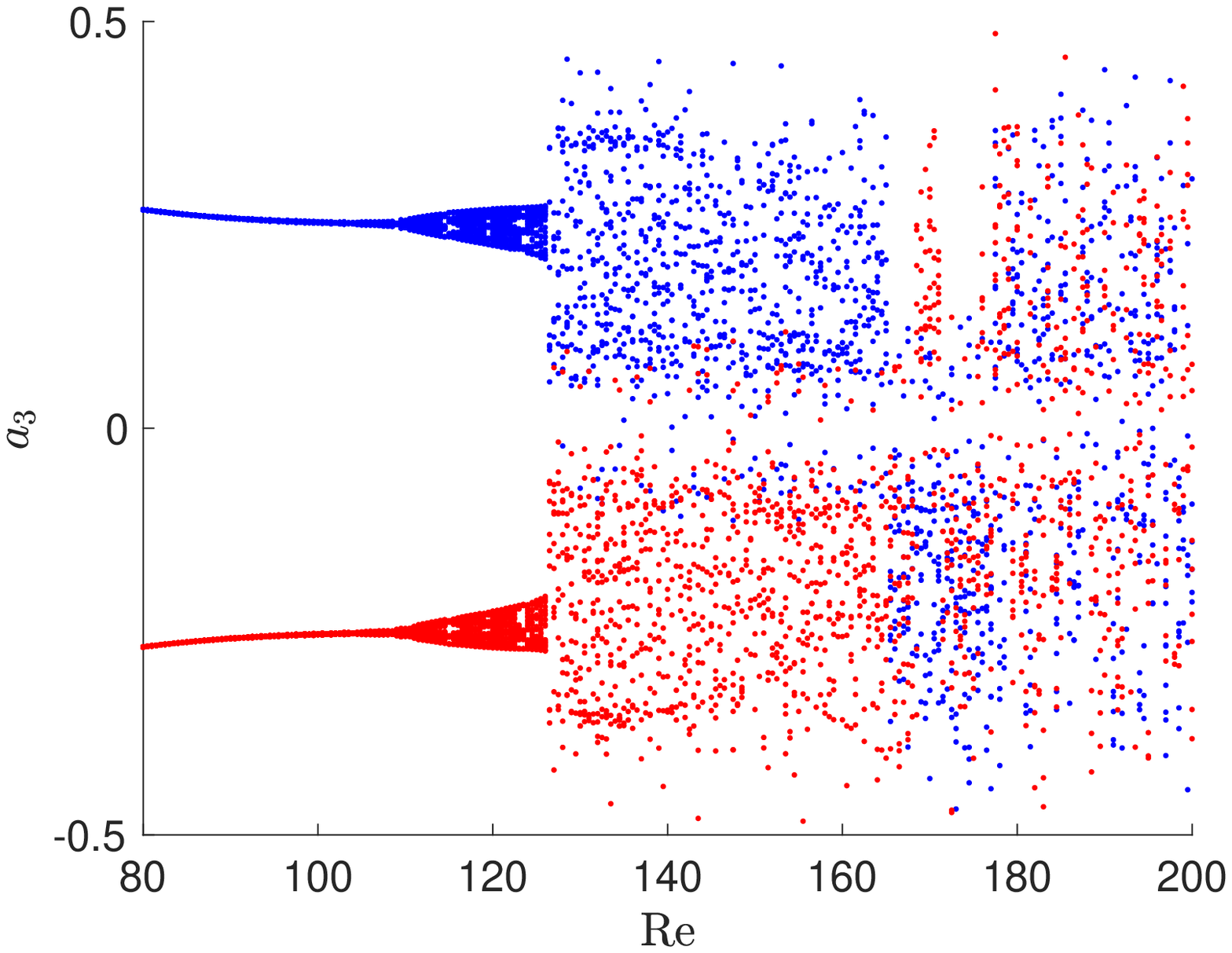}\caption{Poincar\'e section}\end{subfigure}\begin{subfigure}{0.5\textwidth}\includegraphics[width=1.0\textwidth]{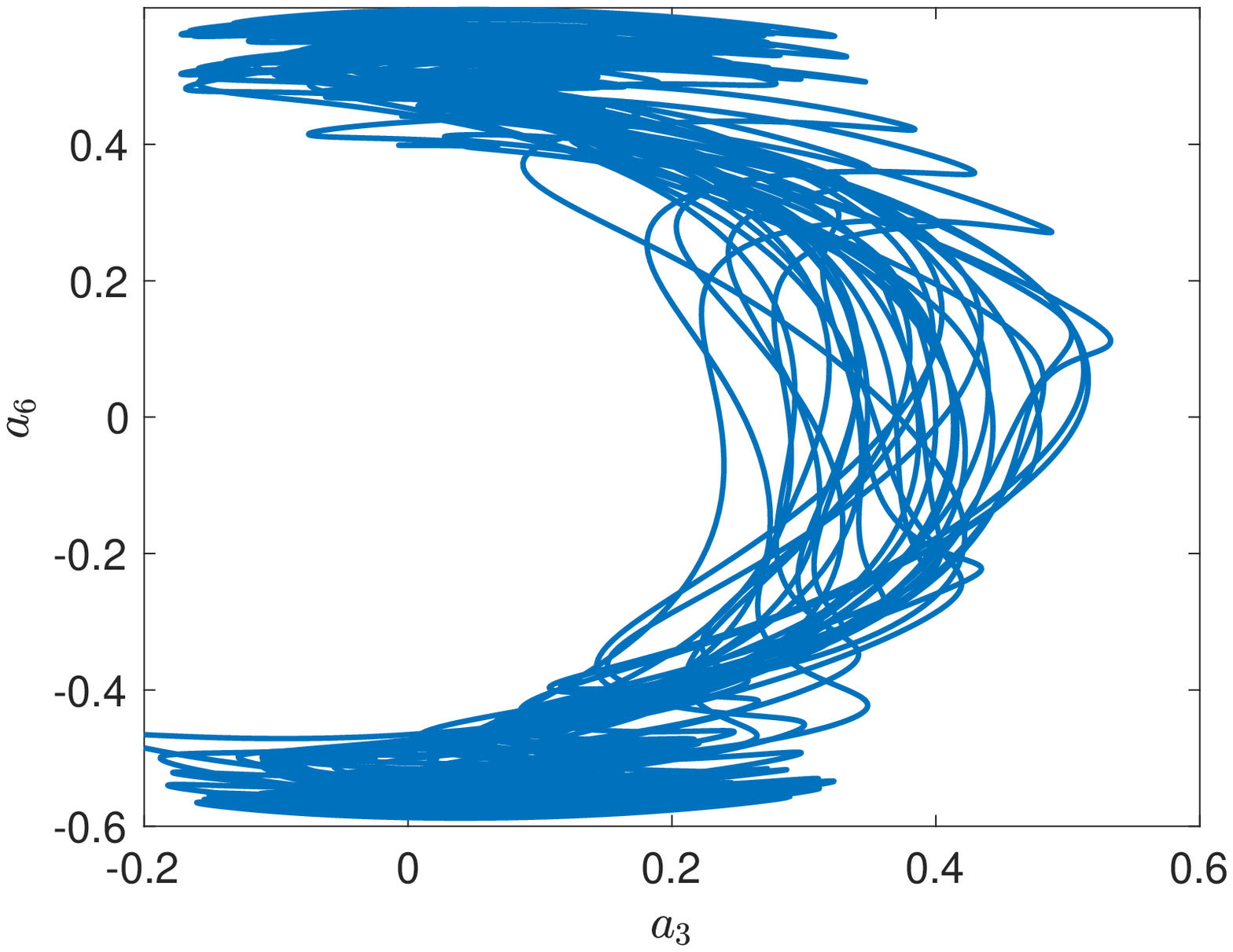}\caption{Phase portrait for $\mathrm{Re}=130$.}\end{subfigure}
\caption{Emergence of a quase-periodic attractor and its collision with the chaotic saddle. Subfigure (a) shows the Poincar\'e section for $P=0.08$. Blue (red) dots show respectively sections tracked in $Re$, starting with the attractor with $a_3>0$ ($a_3<0$). Subfigure (b) shows a phase portrait of the attractor for $\mathrm{Re}=130$.}
\label{fig:poincare_P0.08}
\end{figure}

We now return to the stable limit cycle for $P=0.08$, which is stable for $\mathrm{Re}>72.5$ as seen in section \ref{sec:symbreak}. The Floquet analysis shown in figure \ref{fig:floquet} has a second crossing of the unit circle at $Re=109$, indicating instability. This time, complex conjugate Floquet multipliers leave the unit circle, in a secondary Hopf (or Neimark) bifurcation. This leads to a quasi-periodic attractor, whose behaviour may be seen in the plot of the Poincar\'e section shown in figure \ref{fig:poincare_P0.08}. Notice again that due to the symmetry of the problem, two attractors are present, with either positive or negative values of $a_3$ (only the positive one is shown in figure \ref{fig:symbreak}); each one is tracked separately with increasing $\mathrm{Re}$.

\begin{figure}
\begin{subfigure}{0.5\textwidth}{\includegraphics[width=1.0\textwidth]{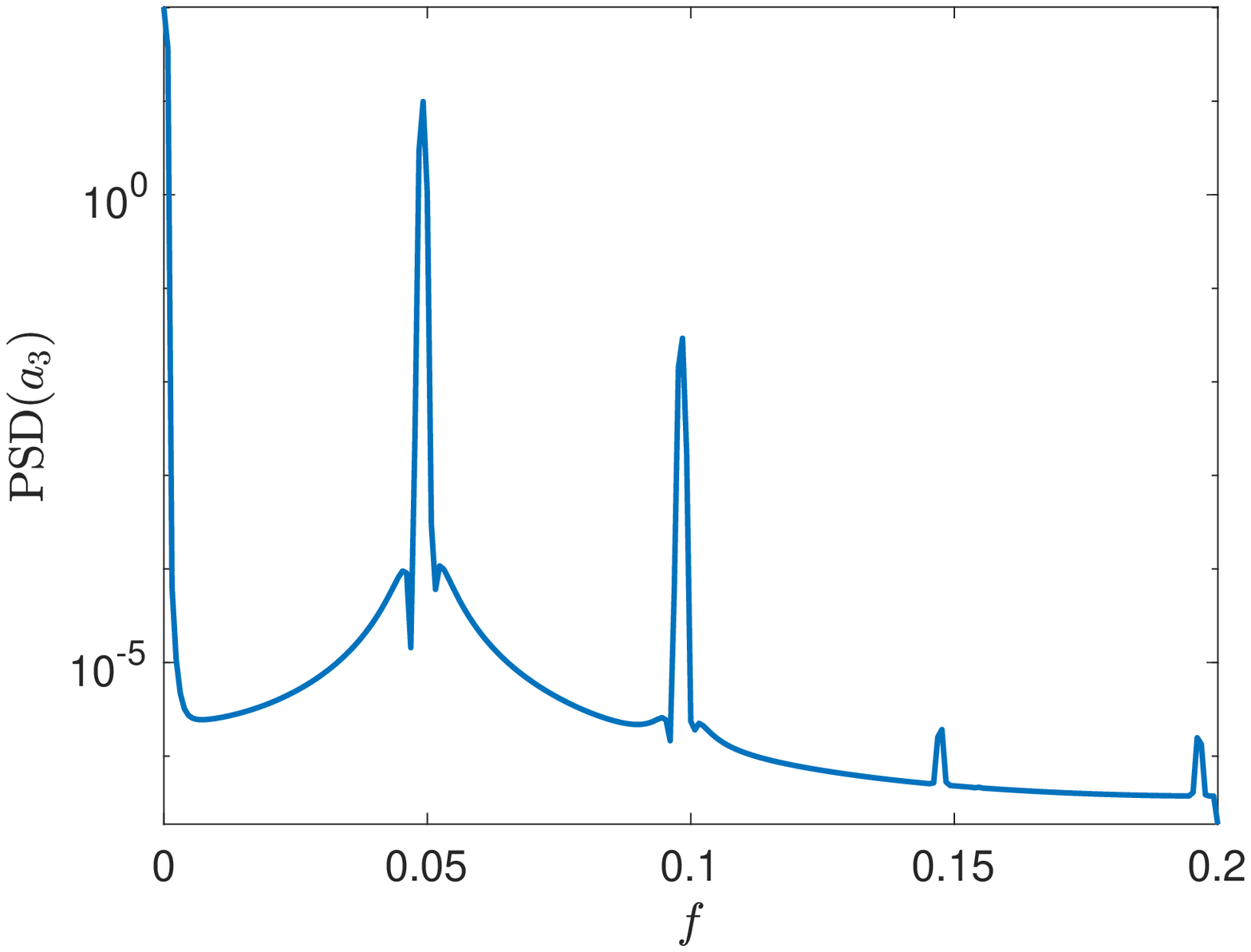}}\caption{$\mathrm{Re}=100$}\end{subfigure}
\begin{subfigure}{0.5\textwidth}{\includegraphics[width=1.0\textwidth]{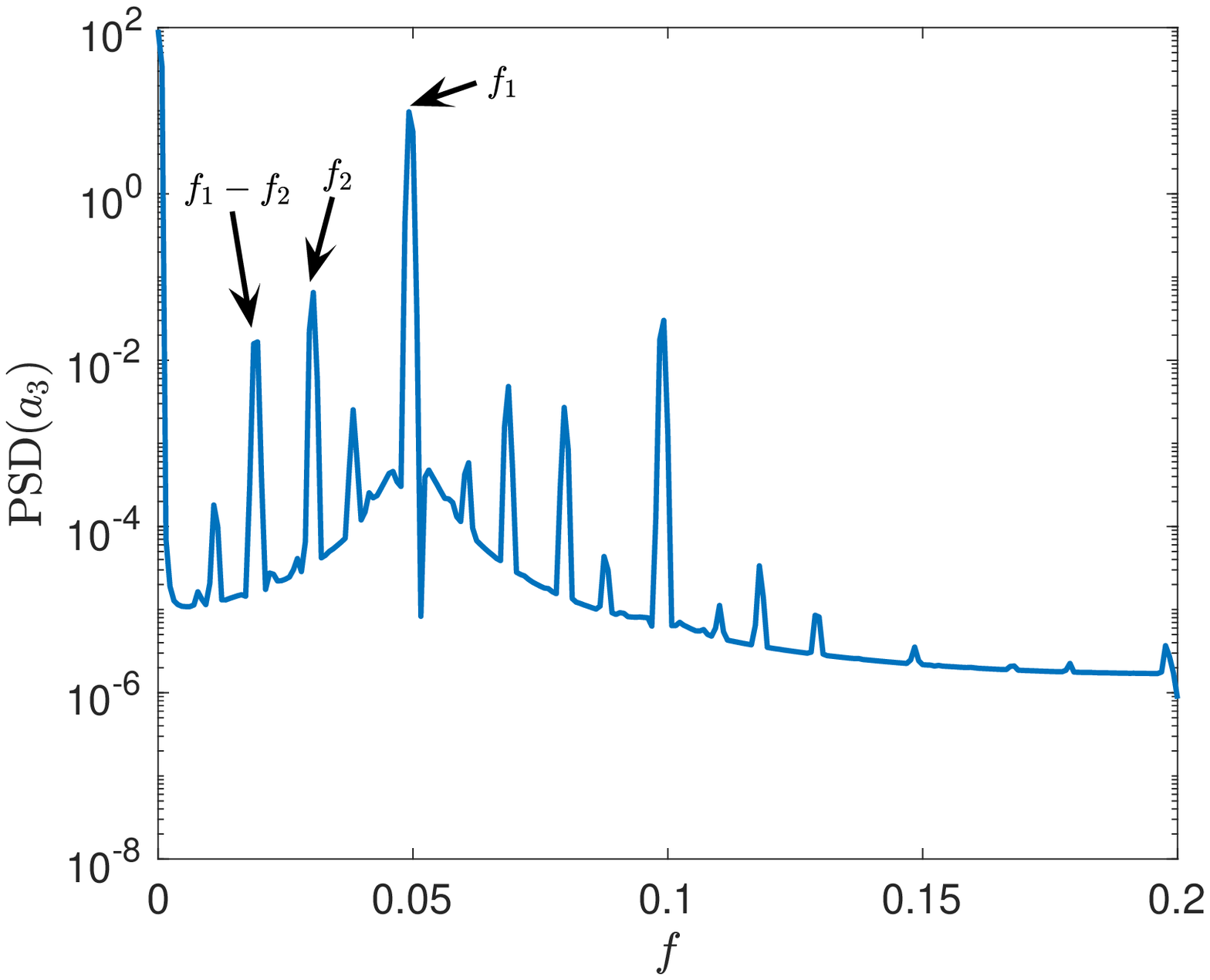}}\caption{$\mathrm{Re}=120$}\end{subfigure}
\caption{Power spectral density of $a_3$ for the attractor, discarding initial transients. Whereas the limit cycle for $\mathrm{Re}=100$ is dominated by $f_1\approx 0.05$ and its harmonics, the quasi-periodic attractor for $\mathrm{Re}=120$ has the appearance of a new frequency $f_2 \approx 0.03$, and the time series has peaks for frequencies given by integer multiples of $f_1$ added to integer multiples of $f_2$}
\label{fig:PSD}
\end{figure}

Figure \ref{fig:poincare_P0.08} shows that at $\mathrm{Re}=109$ the system transitions from a stable period-1 limit cycle to a quasi-periodic attractor, as indicated by the dense region that forms for $109<\mathrm{Re}<126.5$. The appearance of the quasi-periodic attractor is illustrated in figure \ref{fig:PSD}. For $\mathrm{Re}=100$ the attractor is periodic, and the power spectral density (PSD) of figure \ref{fig:PSD}(a) has peaks for a fundamental frequency $f_1\approx 0.05$ and its harmonics. Notice that since the vortex mode $a_3$ was used to calculate the PSD, the dominant frequency corresponds to $2/T$, where $T~\approx 40$ is the period of the cycle shown in figure \ref{fig:floquet}(a), since the oscillations in $a_3$ have a period that is half of the one for the streak mode $a_6$, for instance. Once the attractor becomes quasi-periodic, a second frequency $f_2\approx 0.03$, incommensurate with the previous one, appears in the PSD, as shown in figure \ref{fig:PSD}(b). The time series now has oscillations with these two frequencies, and other ones that correspond to $af_1 + bf_2$ with integer $a, b$. In state space, the trajectories become a torus.

By tracking the Poincar\'e section in figure \ref{fig:poincare_P0.08} we notice a transition of the quasi-periodic behaviour to larger chaotic attractors at $Re=126.5$. Such chaotic attractors are the continuation of the chaotic saddle studied in section \ref{sec:chaoticsaddle}: the sample phase portrait in figure \ref{fig:poincare_P0.08}(b) is close to the chaotic transient in figure \ref{fig:chaotic_saddle_Re80}(a) and the higher-$P$ attractor in figure \ref{fig:chaotic_saddle_Re80}(b). From the observation of the Poincar\'e section, we may conclude that the quasi-periodic attractor collides with the chaotic saddle at $Re=126.5$. It is not straightforward to demonstrate such a collision in a system with more than 3 degrees of freedom, since the Poincar\'e section still has high dimension and allows an apparent crossing of trajectories, which cannot occur in state space. However, due to the resemblance between chaotic attractor at larger $\mathrm{Re}$ and the saddle at lower $\mathrm{Re}$ it is likely that such collision occurs, leading to the disappearance of the quasi-periodic attractor and the emergence of persistent chaos.

\begin{figure}
\begin{subfigure}{0.5\textwidth}{\includegraphics[width=1.0\textwidth]{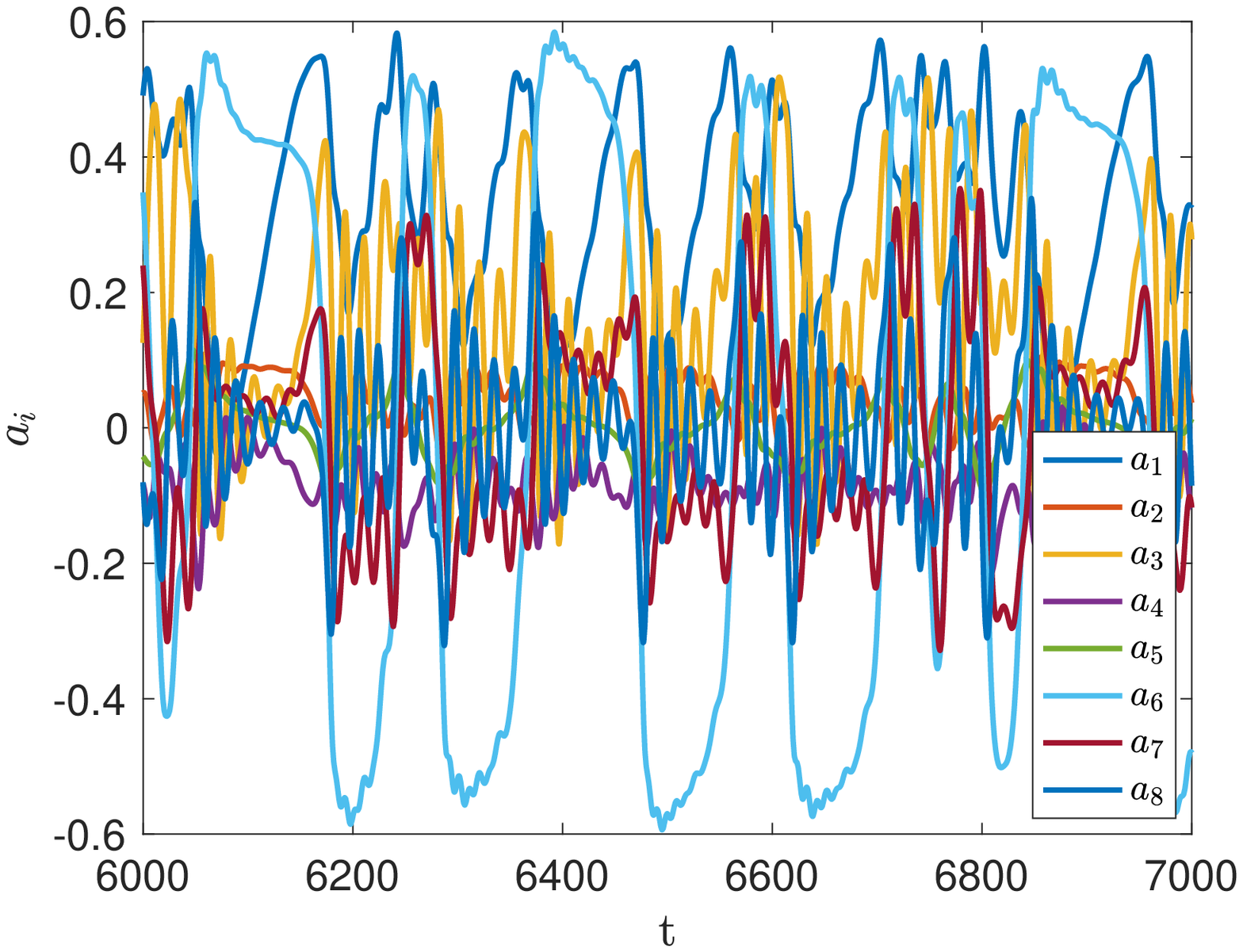}}\caption{Time series}\end{subfigure}
\begin{subfigure}{0.5\textwidth}{\includegraphics[width=1.0\textwidth]{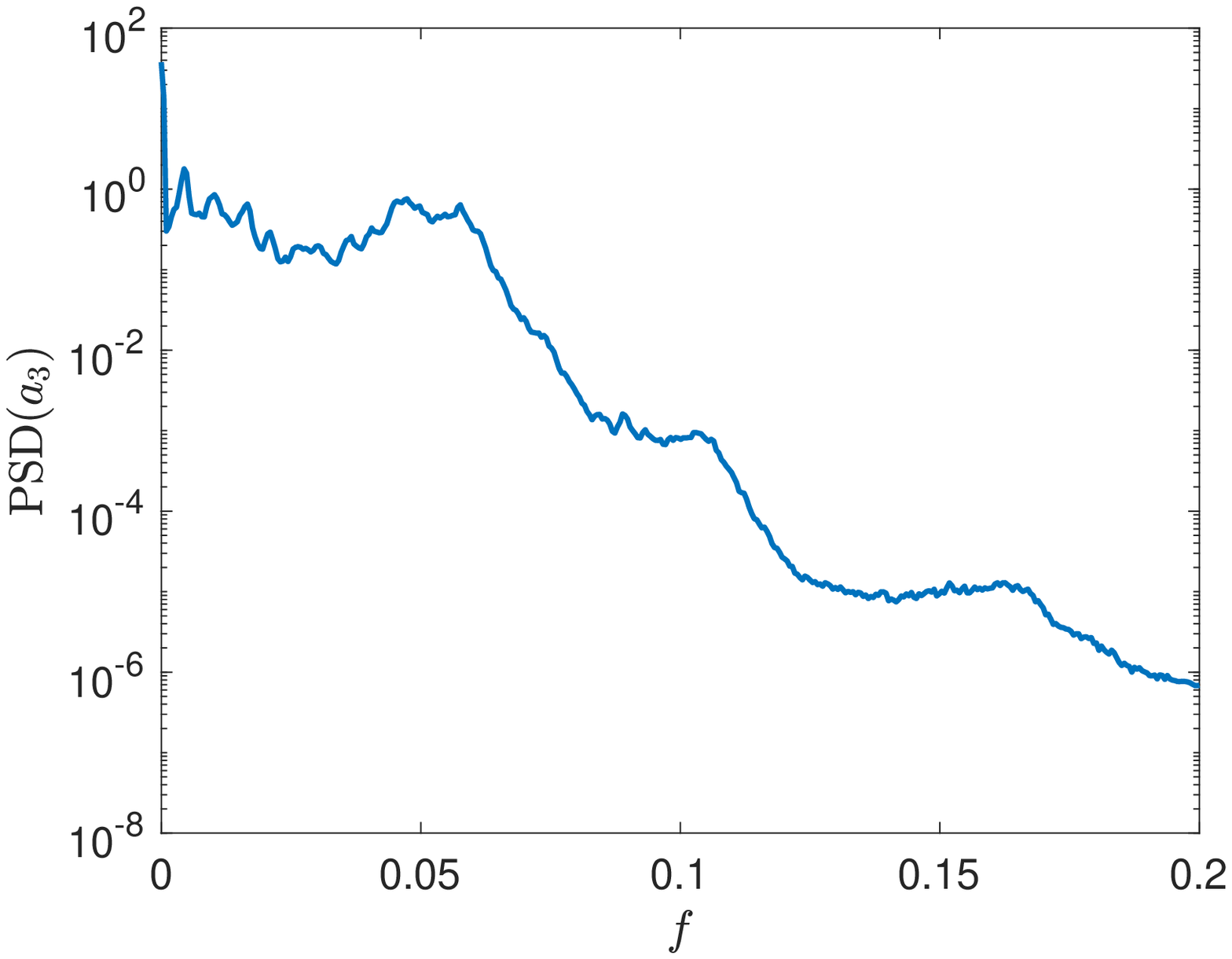}}\caption{Power spectral density of $a_3$}\end{subfigure}
\caption{Results for $\mathrm{Re}=130$.}
\label{fig:PSDRe130}
\end{figure}

We confirm the properties of a chaotic attractor by further inspection of the solutions of the system for $\mathrm{Re}>126.5$. A sample time series for the eight modes is shown in figure \ref{fig:PSDRe130}(a), where no periodicity is evident. The PSD for the KH mode $a_3$ is shown in figure \ref{fig:PSDRe130}(b), showing a broad band behaviour, unlike the spectra of periodic and quasiperiodic attractors in figure \ref{fig:PSD}. The chaotic behaviour is confirmed by a computation of the leading Lyapunov exponent $\lambda_1$ of the system, following \cite{parker2012practical}. This leads to $\lambda_1=0.0186$, a positive value indicating exponential divergence of neighbouring solutions in the attractor, confirming that the system has a marked sensitivity to initial conditions, and is thus chaotic.

Despite the more complex behaviour, the solution has features that are reminiscent of the periodic cycle observed for lower Reynolds numbers. The power spectrum of figure \ref{fig:PSDRe130}(b) shows a broadband peak around $f=0.05$, which is also the peak frequency of the limit cycle shown in figure \ref{fig:PSD}(a). Moreover, inspection of the time series in figure \ref{fig:PSDRe130}(a) shows significant mean-flow distortions (low $a_1$) when streaks ($a_6$) have large amplitude. During these periods vortex modes, in particular $a_3$, remain with low amplitude. The eventual decay of streak amplitude leads to a recovery of the mean shear, and $a_1$ rises to values around 0.5. Once this happens, vortices grow again via the Kelvin-Helmholtz mechanism, until they are quenched by the mean-flow distortion due to a new high-amplitude streak. The amplitude modulation described in section \ref{sec:bifurcations} is now a broad-band, intermittent phenomenon. This can be tentatively compared to observations from turbulent jets, with intermittent events \citep{hileman2005lss,akamine2019conditional} modelled as amplitude modulations of KH wavepackets \citep{cavalieri2011jittering}{, or, alternatively, as intermittent, transient growth of wavepackets \citep{schmidt2019conditional}}. Clearly, care should be taken in drawing such connections between the present simplified ROM and realistic dynamics of high-$\mathrm{Re}$ jets, but further exploration of jet data with the present mechanisms in mind appear to be promising.

\subsection{Merging crisis}
\label{sec:mergingcrisis}

\begin{figure}
\begin{subfigure}{0.5\textwidth}\includegraphics[width=1.0\textwidth]{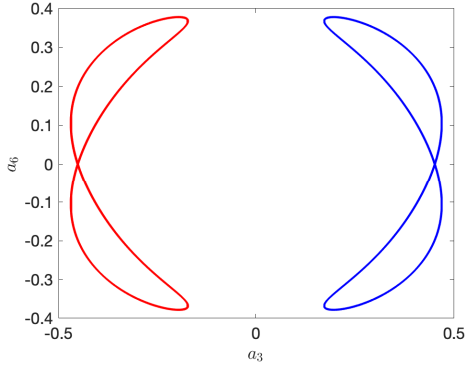}\caption{$Re=100$, periodic solution}\end{subfigure}\begin{subfigure}{0.5\textwidth}\includegraphics[width=1.0\textwidth]{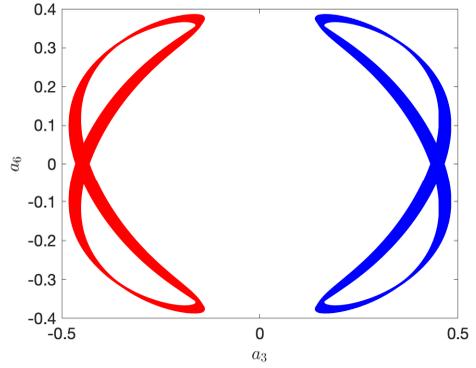}\caption{$Re=120$, quasi-periodic solution}\end{subfigure}
\begin{subfigure}{0.5\textwidth}\includegraphics[width=1.0\textwidth]{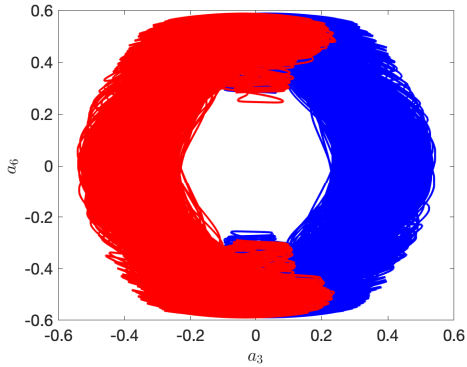}\caption{$Re=150$, chaotic attractor ($\lambda_1=0.0186$)}\end{subfigure}\begin{subfigure}{0.5\textwidth}\includegraphics[width=1.0\textwidth]{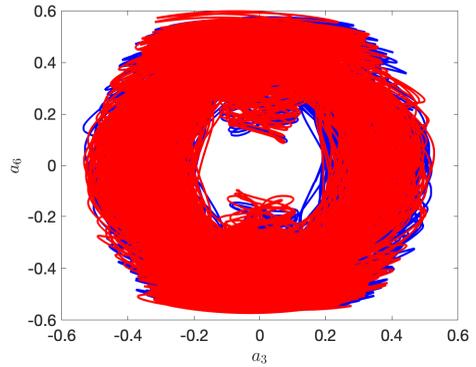}\caption{$Re=180$, merged chaotic attractor ($\lambda_1=0.0202$)}\end{subfigure}
\caption{Phase portraits of attractors of the $P=0.08$ system at selected Reynolds numbers. Blue and red lines show attractors with positive and negative $a_3$, respectively. Notice that for $Re=180$ the attractors of lower $Re$ are already merged, thus blue and red lines refer to the same set. For the chaotic attractors, $\lambda_1$ is the largest Lyapunov exponent.}
\label{fig:P0.08_attractors}
\end{figure}

\begin{figure}
\includegraphics[width=1.0\textwidth]{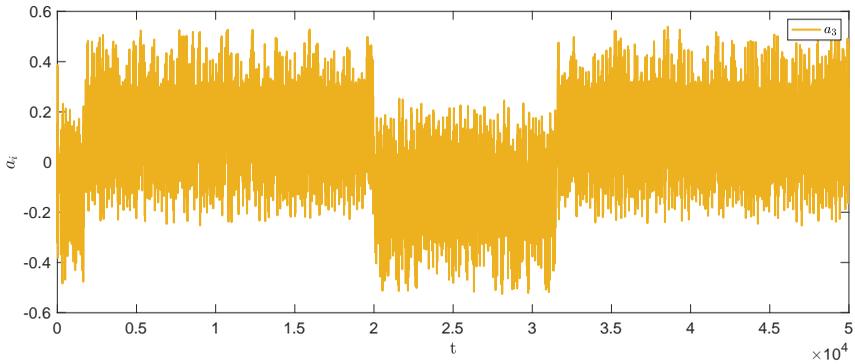}
\caption{Time series of mode 3 for $\mathrm{Re}=180$.}
\label{fig:a3mergingcrisis}
\end{figure}

The Poincar\'e plot in figure \ref{fig:poincare_P0.08} shows separate attractors up to $\mathrm{Re}=164.5$, represented with blue and red dots. Thus, for the periodic, quasiperiodic and chaotic attractors of the system the dynamics remain confined to either positive or negative $a_3$. The time series in  figure \ref{fig:PSDRe130}(a), for instance, shows the evolution of the system for the attractor with mostly $a_3>0$.  Sample phase portraits of the various regimes are shown in figure \ref{fig:P0.08_attractors}, which displays both positive and negative periodic (a), quasiperiodic (b) and chaotic (c) attractors. For such Reynolds numbers, the system stays indefinitely at one of the attractors, and no numerical integration performed here showed any sign of switch. However, at $\mathrm{Re}=164.5$ the chaotic attractors undergo a merging crisis, similar to what was seen in section \ref{sec:chaoticsaddle}. The present crisis leads to a larger attractor spanning both positive and negative $a_3$. The system spends extended periods with $a_3$ mostly of positive sign, and abruptly switches to mostly negative $a_3$. This is exemplified in the time series shown in figure \ref{fig:a3mergingcrisis}, obtained for $\mathrm{Re}=180$.

\begin{figure}
\centerline{\includegraphics[width=0.5\textwidth]{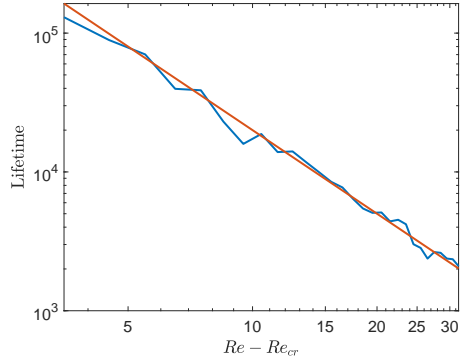}}
\caption{Lifetime of trajectories in an assymmetric chaotic saddle after the merging crisis at $\mathrm{Re}_{cr}=164.5$. The red line indicates a $(\mathrm{Re}-\mathrm{Re}_{cr})^{-2}$ dependence}
\label{fig:lifetimes_merging_crisis}
\end{figure}

The intepretation of the results of figure \ref{fig:a3mergingcrisis} is that the time series of the merged chaotic attractor comprises extended periods where the system stays in a trajectory resembling one of the lower-$Re$ asymmetric chaotic attractors, as discussed by \cite{grebogi1987critical}. In this case, close to the critical Reynolds number $\mathrm{Re}_{cr}$ of the crisis, the typical lifetime of the solution at one of the original attractors, before switching to the other one, has an expected power-law behaviour $(\mathrm{Re}-\mathrm{Re}_{cr})^{\gamma}$, with $\gamma$ being a negative exponent. The lifetime statistics were obtained close to $\mathrm{Re}_{cr}=164.8$, and are shown in figure \ref{fig:lifetimes_merging_crisis}. The lifetime closely follows a $(\mathrm{Re}-\mathrm{Re}_{cr})^{-2}$ dependence and supports the conclusion that a merging crisis occurs for this Reynolds number.

Past the merging crisis, the evolution of the KH mode $a_3$ now involves modulation in amplitude and changes in phase. In the present model all modes are fixed at a given streamwise and spanwise position. As a single Fourier mode was included for each considered wavenumber, it is impossible to represent travelling waves or relative periodic orbits, and the only phases in $a_3$ are 0 or $\pi$; alternatively, one may think of vortex positions at $x=0$ or $x=L_x/2$, and each jump in the merged chaotic attractor represents a phase jump in $a_3$. A more complete model would include two modes for each wavenumber, allowing a continuous change of phase. In this case, the dynamics of the larger, merged attractor would lead to continuous phase drifts of the vortices. This would represent a varying phase speed in addition to amplitude modulation. 

\begin{figure}
\centerline{\begin{subfigure}{0.9\textwidth}\includegraphics[width=1.0\textwidth]{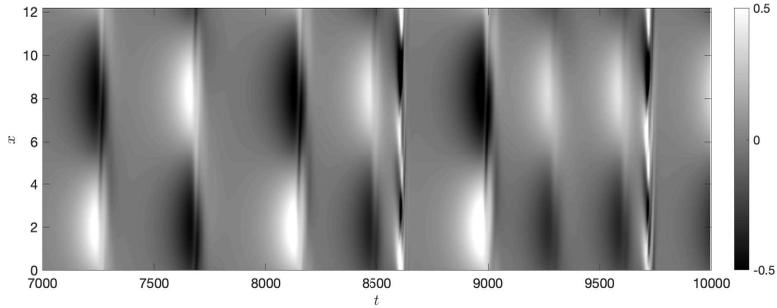}
\caption{DNS results, with the same configuration of NC, but at $Re=205$}
\end{subfigure}}
\centerline{\begin{subfigure}{0.9\textwidth}\includegraphics[width=1.0\textwidth]{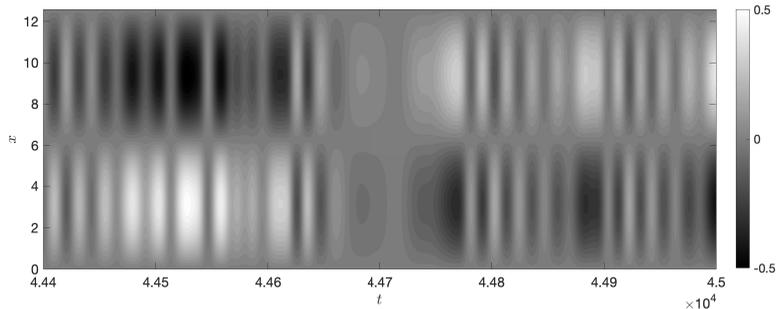}\caption{ROM at $Re=200$}\end{subfigure}}
\caption{Vortices in the NC DNS for $\mathrm{Re}=205$, and in the present model for $\mathrm{Re}=200$, illustrated by fluctuations of the streamwise velocity on $y=1$ for the ROM and for $y=0.5$ for the DNS.}
\label{fig:NC_Re205}
\end{figure}

This feature can also be seen by increasing the Reynolds number of the DNS. Streamwise velocity fields at $y=1.5141$ were extracted from a simulation at $Re=205$ (keeping the same numerics as in NC), and these are shown in figure \ref{fig:NC_Re205}(a). The streamwise averaged field was subtracted from the velocity to highlight the presence of the KH vortices. Figure \ref{fig:NC_Re205} shows that, differently from the $Re=200$ case reported in NC, the dynamics of the flow for this Reynolds number is no longer periodic. In this case, vortices suffer a change in phase in some times, similar to the present model, {with space-time behaviour of vortices reconstructed using $a_3$ and $\mathbf{u}_3$, taken at $y=1$. As in previous comparisons between ROM and DNS, quantitative agreement is not expected due to the non-slip conditions in the DNS, but the results show that both the DNS and ROM display phase jumps of $\pi$ at specific times.}

An association of this phenomenon to jet dynamics would have important consequences. Changes in phase speed greatly affect sound radiation by wavepackets in subsonic flows, as the dependence on phase speed is exponential~\citep{crighton1975bpa,cavalieri2011jittering,koenig2016jet}. An intermittent increase of phase speed would lead to a burst in sound radiation. The observation of space-time measurements of turbulent jets suggests that instantaneous changes in phase speed may occur~\citep{breakey2017experimental}, but to the best of our knowledge this has not been properly studied. Again, one should bear in mind that the observed trends come from a simplified representation of a parallel shear layer, but the present observations suggest directions for the investigation of data from jets, which may uncover the mechanisms at hand.

{In summary, by studying the reduced-order model with increasing Reynolds number we obtain progressively richer dynamics, including limit cycles, quasi-periodic attractors, chaotic saddles and attractors. The low dimension of the model, involving eight modes, limits the dynamics to a minimal representation. Despite the low order of the system, the dynamics resemble observations of turbulent jets. The non-linear interactions included in the present low-order system, shown in eq. (\ref{eq:ROM}), point thus to phenomena of potential relevance in flows of interest. Further investigation of the non-linear interactions in turbulent jets, using appropriate signal processing \citep{schmidt2020bispectral}, is a promising direction, and the present reduced-order model, as well as its solutions, may aid in the interpretation of data. The reduced number of non-linear interactions in the model may also help in the development of non-linear models for spatially developing shear layer and jets \citep{suponitsky2010linear,wu2016nonlinear,zhang2020nonlinear}.}



\section{Conclusions}
\label{sec:conclusions}

The non-linear dynamics of a shear layer are studied using a reduced-order model. A shear layer confined between parallel walls and driven by a body force, introduced by \cite{nogueira2021dynamics} (NC), prevents the spatial development of the flow and thus allows the use of periodic boundary conditions in streamwise and spanwise directions, which greatly simplifies the problem. The same configuration is considered here, but instead of the non-slip boundary conditions used in the DNS in NC, we adopt free-slip boundary conditions on the parallel walls. This ensures that the confined shear layer does not develop near-wall fluctuations that would clearly be absent in free shear layers and jets. Moreover, by using free-slip boundary conditions, a reduced-order model may be obtained using Fourier modes in all three spatial directions, and application of a Galerkin projection of the Navier-Stokes equation leads to a system of ordinary differential equations, similar to related works on wall-bounded turbulence \citep{waleffe1997self,moehlis2004low,cavalieri2021structure}. 

The procedure aims to build and study a simplified system in order to reveal non-linear dynamics of shear layers, which may shed light on the behaviour of spatially developing flows such as jets. Turbulent jets have been thoroughly studied using linearised models and input/output analysis in recent years~\citep{garnaud2013preferred,jeun2016input,schmidt2018spectral,lesshafft2019resolvent}, but features such as amplitude and phase modulation of Kelvin-Helmholtz wavepackets, referred to as ``jitter'', are more challenging to model using linear theory. Jitter in amplitude and phase has been observed in numerical and experimental data and related to the sound radiation, with particular relevance for subsonic flows~\citep{cavalieri2011jittering,baqui2015coherence,cavalieri2019wave}. The analysis in NC used a direct numerical simulation of a confined shear layer, showing that non-linear dynamics lead to the appearance of a stable limit cycle with amplitude modulation of Kelvin-Helmholtz vortices, in a behaviour reminiscent of amplitude jitter seen in jet data. This direction is further pursued here by the construction of a reduced-order model (ROM) with the dominant structures studied in NC, which allows a more straightforward analysis of the non-linear dynamics of the confined shear-layer configuration.

Eight modes were selected to build the ROM, based on the observations in NC. Such modes represent the mean flow, Kelvin-Helmholtz vortices, rolls, streaks, and oblique waves. For the mean flow, vortices and oblique waves, a pair of Fourier modes were needed for each structure in order to ensure a consistent representation of the dynamics. The body force that drives the shear layer was taken as a sum of two Fourier modes, with the second parametrised by a forcing parameter $P$ that is related to the growth rate of the Kelvin-Helmholtz instability of the laminar solution. The non-linear dynamics of the resulting ROM could be analysed in some detail, thanks to the reduced computational cost to obtain numerical solutions and their stability analysis. This provides a picture of how the confined shear layer transitions to the limit cycle studied in NC, and how such cycle eventually transitions to chaotic behaviour as the Reynolds number is increased.

The first bifurcations identified for the ROM are a pitchfork leading to saturated vortex solutions, and a Hopf bifurcation leading to a stable limit cycle. This limit cycle resembles the observations in NC, with amplitude modulation of vortices related to oscillations in rolls, streaks and oblique waves. As the Reynolds number is increased, we observe the emergence of a chaotic saddle. This was studied by varying the forcing parameter $P$, showing the emergence of chaotic attractors via a period doubling cascade as $P$ is increased. {Such attractors eventually merge and form a larger chaotic attractor for $P=0.09$. At $Re=80$ and $P=0.08$ such attractor loses stability and becomes a chaotic saddle, with transient chaos.} Thus, most initial disturbances for $P=0.08$ and Reynolds number between 72.5 and 109 display transient chaotic behaviour which eventually settles into the stable limit cycle that emerged from the first Hopf bifurcation. Since the reduced-order model has symmetries, with corresponding solutions obtained by considering shifts in streamwise or spanwise direction, all such solutions appear in pairs, with one of them corresponding to a vortex mode $a_3>0$, and the other related to $a_3<0$. Such solutions display time-periodic (for the limit cycle) or aperiodic (for the chaotic saddle) modulation of vortex amplitude, a behaviour that resembles the jitter in amplitude observed in high Reynolds number jets.

A further increase of $Re$ leads to the appearance of a quasi-periodic attractor at $Re=109$. This attractor collides with the chaotic saddle at $Re=126.5$, leading to the appearance of a chaotic attractor. The time series of the various modes show broadband behaviour, with no periodicity, in a more complex behaviour that approaches what is seen in turbulence. The symmetries of the system again imply that two chaotic attractors are formed, with positive or negative vortex time coefficient $a_3$. Such attractors suffer a merging crisis at $Re=164.8$, and become a larger chaotic attractor where solutions spend some time at one of the original attractors and then ``jump'' to the other one. This implies that the solution starts to display jumps between positive and negative values of the vortex $a_3$, which in turn means that vortices have a $\pi$ phase jump in $x$. Such phase jitter is again reminiscent of observations of Kelvin-Helmholtz wavepackets in turbulent jets. Thus, the present ROM may be thought as a minimal model that mimics jitter in amplitude and phase of Kelvin-Helmholtz vortices. Its modes and their interaction are minimal ingredients to obtain dynamics that resemble observations of high-$Re$ jets.

Keeping in mind that several simplifications were applied to both the flow configuration (by considering a confined shear layer) and the underlying equations (by reducing the model using a Galerkin projection with eight modes), the present reduced-order model opens a number of interesting directions for the study of turbulent free-shear flows. The ROM has closed-form equations that include non-linear interactions between modes that have turbulent jet counterparts, such as Kelvin-Helmholtz vortices~\citep{cavalieri2013wavepackets}, rolls and streaks~\citep{nogueira2019large}. By studying the non-linear interactions among the various structures in the model, one may better understand how the corresponding modes interact in turbulent jets, which is not possible using standard linear input/output analysis. The reduced-order model displays clear correlations between vortices and streaks, which may be further investigated using dedicated signal processing of numerical or experimental databases.

It is clear that high-Reynolds number jets will display dynamics that are not comprised by the model. However, it is hoped that the rich non-linear dynamics displayed by the ROM may shed light on behaviours of more complex flows. In the past, reduced-order models of wall-bounded turbulence by \cite{waleffe1997self} and \cite{moehlis2004low} helped to understand the relationship of rolls, streaks and sinuous disturbances, and showed that under some conditions turbulence in such configurations had finite lifetimes, an observation later confirmed by the experiments of \cite{hof2006finite}. Reduced-order models were important in defining the concept of edge states between laminar and turbulent solutions \citep{skufca2006edge}, a concept later extended to wall-bounded flows using full direct numerical simulations~\citep{schneider2007turbulence,duguet2013minimal}. It is hoped that the present reduced-order model will also enable a number of approaches based on dynamical systems theory, which may help probing the non-linear dynamics of turbulent shear layers and jets.

\section*{Acknowledgments}

This work was supported by FAPESP Grant No. 2019/27655-3 and CNPq Grant Nos. 313225/2020-6 and 306920/2020-4. A numerical implementation of the present reduced-order models is available by request to the first author.

Declaration of Interests. The authors report no conflict of interest.

\appendix

\section{Saddle-node bifurcations of asymmetric limit cycles}

\begin{figure}
\begin{subfigure}{0.5\textwidth}{\includegraphics[width=1.0\textwidth]{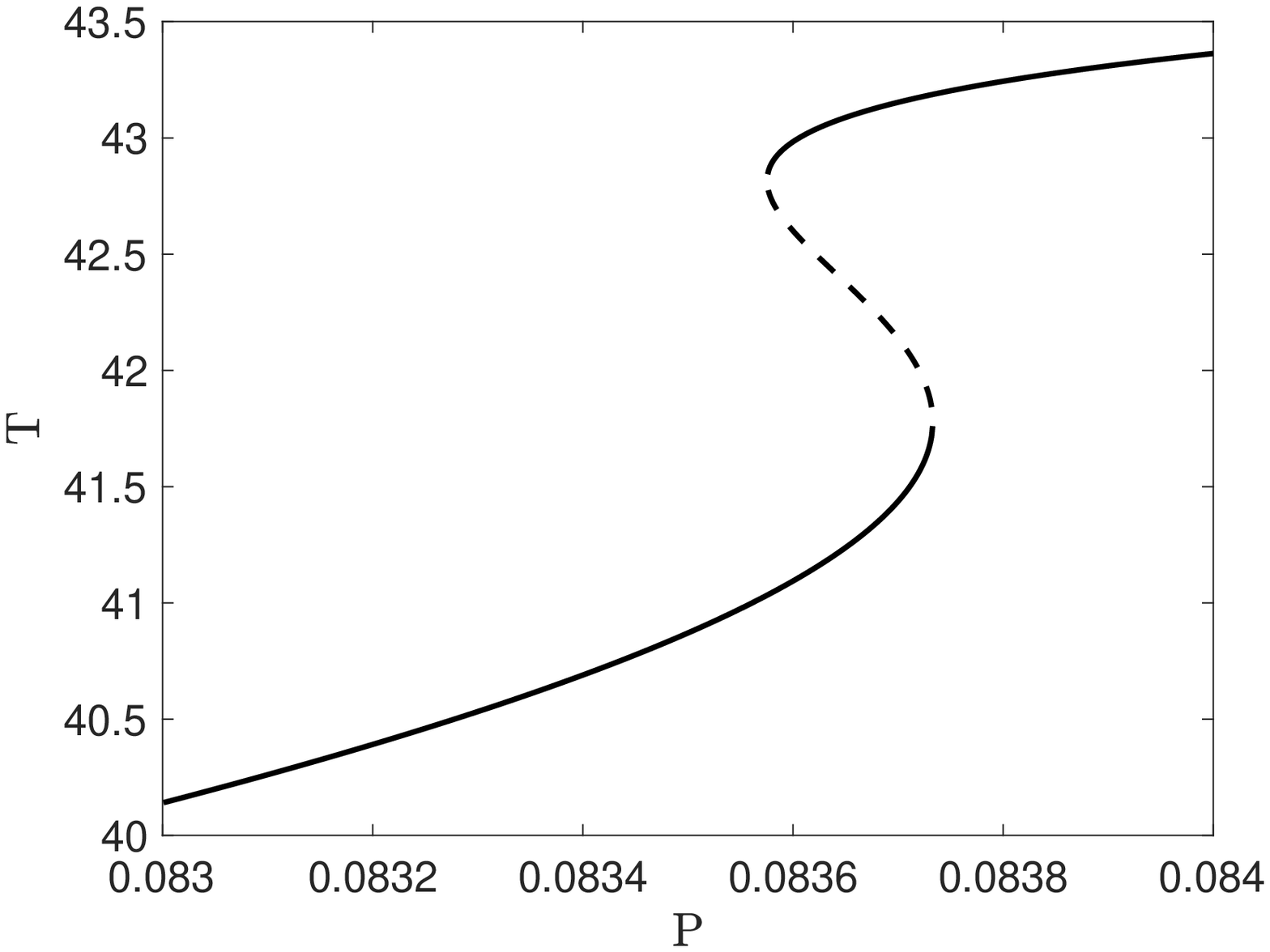}}\end{subfigure}
\begin{subfigure}{0.5\textwidth}{\includegraphics[width=1.0\textwidth]{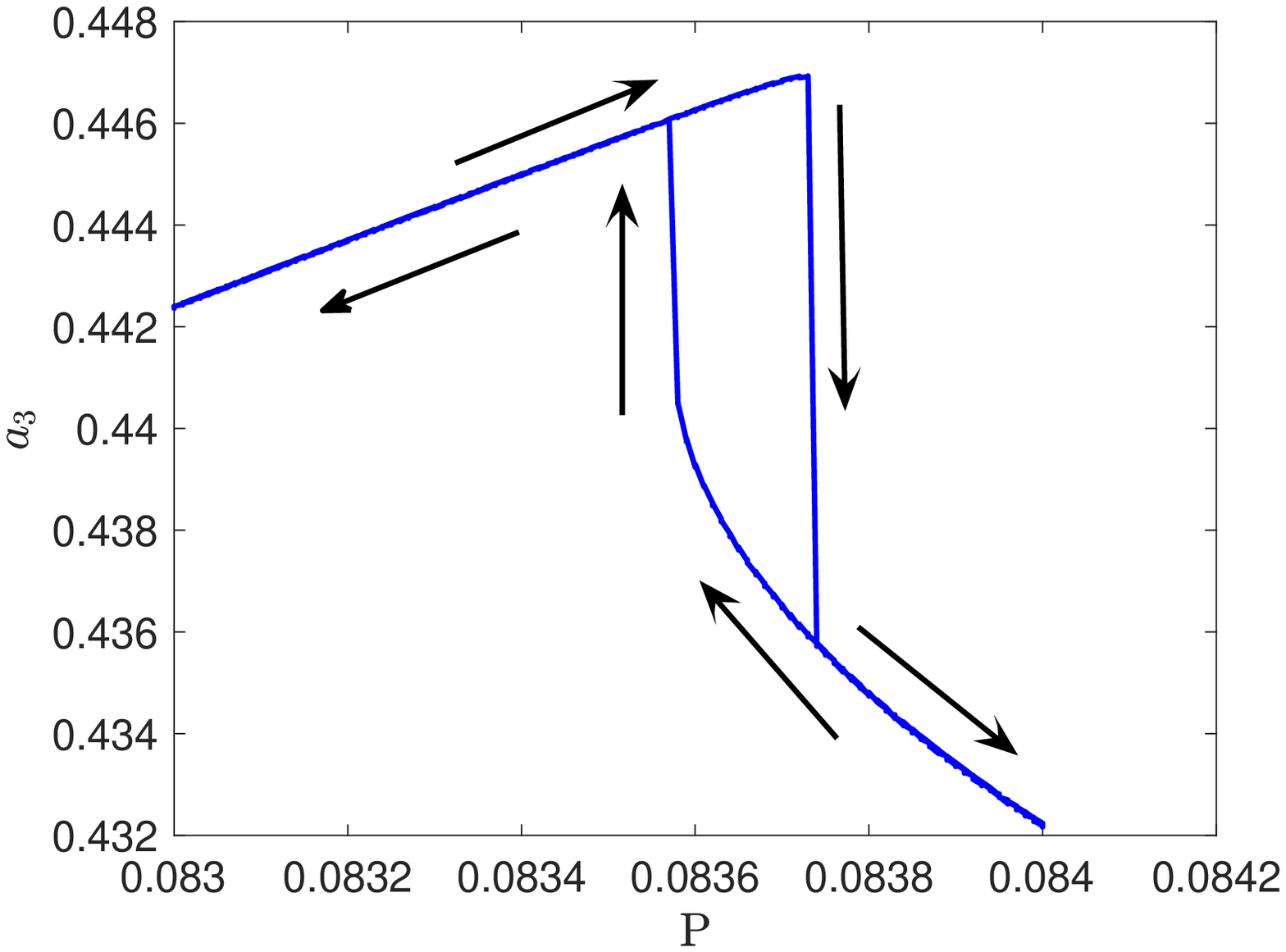}}\end{subfigure}\caption{Saddle-node bifurcations for $\mathrm{Re}=54$: (a) period of periodic solutions, with full lines showing stable cycles and the dashed line showing the unstable one, and (b) Poincar\'e section at $a_5=0$ and $\dot{a}_5<0$, with arrows highlighting the path in the plot as $P$ is increased and then decreased.}
\label{fig:hysteresis}
\end{figure}

This Appendix briefly studies the saddle-node bifurcations of the asymmetric limit cycles studied in section \ref{sec:symbreak}. Figure \ref{fig:hysteresis}(a) shows periodic solutions with $\mathrm{Re}=54$, and $P$ between 0.083 and 0.084. There are two stable limit cycles, shown with full lines, that are connected to an unstable cycle by saddle-node bifurcations for $P=0.08373$ and $P=0.08358$. This pair of saddle-node bifurcations leads to hysteresis, as shown in figure \ref{fig:hysteresis}(b): as $P$ is increased above 0.083, the system remains in the lower-branch cycle until the first saddle-node bifurcation, where it jumps to the upper-branch solution. As $P$ is decreased from 0.084, the system remains in the upper branch until it reaches the second saddle-node bifurcation, where the solution jumps to the lower branch. This explains the jump observed in figure \ref{fig:symbreak}.

\bibliographystyle{abbrvjfm}


\end{document}